\documentclass[structabstract]{aa}  
\usepackage{graphicx}
\usepackage{txfonts}
\usepackage{natbib}
\usepackage{lscape}

\begin{document}

\title{Polycyclic aromatic hydrocarbon (PAH) luminous galaxies at $z\sim 1$}

   \author{T. Takagi \inst{1}
                    \and
          Y. Ohyama \inst{2}  \and
	T. Goto \inst{3,4}  \and
	H. Matsuhara \inst{1} \and
	S. Oyabu \inst{1} \and
	T. Wada \inst{1} \and	
          C.P. Pearson \inst{5} \and		
	H.M. Lee \inst{6} \and 
	M. Im \inst{6} \and 
	M.G. Lee \inst{6} \and 
	H. Shim  \inst{6} \and 
	H. Hanami \inst{7} \and 
	T. Ishigaki  \inst{8} \and 
	K. Imai \inst{9} \and 
	G.J. White \inst{5,10} \and 
	S. Serjeant \inst{10} \and 
	M. Malkan \inst{11} 
          }

   \institute{
   Institute of Space and Astronautical Science, Japan Aerospace Exploration Agency, 
   Sagamihara, Kanagawa 229-8510, Japan \\
                   \email{takagi@ir.isas.jaxa.jp}
         \and
             Academia Sinica, Institute of Astronomy and Astrophysics, Taiwan 
             \and
             Institute for Astronomy, University of Hawaii, 2680 Woodlawn Drive, Honolulu, HI, 96822, USA
             \and
             National Astronomical Observatory, 2-21-1 Osawa, Mitaka, Tokyo, 181-8588, Japan
             \and
             Rutherford Appleton Laboratory, Chilton, Didcot, Oxfordshire, OX11 0QX, UK 
             \and 
             Department of Physics and Astronomy, FPRD, Seoul National University, Shilim-Dong, Kwanak-Gu, 
             Seoul 151-742, Korea
             \and
             Physics Section, Faculty of Humanities and Social Sciences, Iwate University, Morioka 020-8550, Japan
             \and
             Asahikawa National College of Technology, Asahikawa, Hokkaido 071-8124, Japan
             \and
             TOME R\&D Inc. Kawasaki, Kanagawa 213-0012, Japan
             \and 
             Astrophysics Group, Department of Physics, The Open University, Milton Keynes, MK7 6AA, UK
             \and
             Department of Physics and Astronomy, UCLA, Los Angeles, CA, USA
             }

   \date{Received 14 October 2009; accepted }

 
  \abstract
   {}
   {Using an AKARI multi-wavelength mid-infrared (IR) survey, we identify luminous starburst galaxies at $z\ga 0.5$ based on the PAH luminosity, and investigate the nature of these PAH-selected starbursts.}
   {  An extragalactic survey with AKARI towards the north ecliptic pole (NEP), the NEP-Deep survey, is unique in terms of a comprehensive wavelength coverage from 2 to 24\,$\mu$m using all 9 photometric bands of the InfraRed Camera (IRC). This survey allows us to photometrically identify galaxies whose mid-IR emission is clearly dominated by PAHs. We propose a single colour selection method to identify such galaxies, using two mid-IR flux ratios at 11-to-7\,$\mu$m and 15-to-9\,$\mu$m (PAH-to-continuum flux ratio in the rest-frame), which are useful to identify starburst galaxies at $z\sim 0.5$ and 1, respectively. We perform a fitting of the spectral energy distributions (SEDs) from optical to mid-IR wavelengths, using an evolutionary starburst model with a proper treatment of radiative transfer (SBURT), in order to investigate their nature. 
   }
   { The SBURT model reproduces observed optical-to-mid-IR SEDs of more than a half of PAH-selected galaxies. 
   Based on the 8\,$\mu$m luminosity, we find ultra luminous infrared galaxies (ULIRGs) among PAH-selected galaxies. Their PAH luminosity is higher than local ULIRGs with a similar luminosity, and the PAH-to-total IR luminosity ratio is consistent with that of less luminous starburst galaxies. They are a unique galaxy population at high redshifts and we call these PAH-selected ULIRGs ``PAH-luminous'' galaxies. Although they are not as massive as submillimetre galaxies at $z\sim 2$,  
   they have the stellar mass of $>3\times 10^{10}$\,M$_\odot$ and therefore moderately massive. 
   }
   {}


   \maketitle
%

\section{Introduction}

In recent studies of galaxy formation and evolution, the most massive and luminous galaxies have played a 
leading role. Most luminous galaxies, such as submillimetre galaxies and ultraluminous infrared galaxies (ULIRGs\footnote{Galaxies with a total infrared (8 -- 1000\,$\mu$m) luminosity of 10$^{12\mathrm{-}13}$\,L\,$_\odot$.}) 
at high redshifts are now believed to be the progenitors of massive spheroids, and highlight the early formation 
of massive galaxies \citep[e.g.][]{2004ApJ...616...71S,2005ApJ...622..772C,2008ApJ...680..246T}. This supports the scenario of anti-hierarchical or `down-sizing' galaxy formation \citep[e.g.][]{2004Natur.428..625H,2006A&A...453L..29C}. 

Mid-infrared (IR) surveys are a powerful tool for studying the most luminous galaxies at high redshifts, by 
providing the infrared luminosity function and the star formation rate in a large 
cosmic volume as a function of redshift \citep[e.g][]{2005ApJ...630...82P,2005ApJ...632..169L,2006MNRAS.370.1159B,2007ApJ...660...97C,Goto_LF}. 
However, this work can suffer uncertainty from the $k$-correction, which requires knowing the 
appropriate mid-IR spectral energy 
distribution (SED). 
In fact, sensitive IRS observations with the Spitzer Space Telescope reveal that the mid-IR spectra of submillimetre galaxies resembles that of over 2 orders of magnitude less luminous starburst galaxies  \citep{2007ApJ...660.1060V,2008ApJ...675.1171P,2008ApJ...677..957F},  
which have prominent emission features of polycyclic aromatic hydrocarbons (PAHs).
This raises questions about  the nature of most luminous galaxies, and how they differ systematically from low-$z$ to high-$z$. 

Multi-wavelength mid-IR surveys with AKARI provide a unique
measure of the mid-IR properties of infrared galaxies at high redshifts with a statistically valid sample. 
\cite{2007PASJ...59S.557T} demonstrate that AKARI/IRC all-band photometry is capable of 
identifying the approximate spectral shape of the PAH emission, specifically the steep rise of flux at 
the blue side of the PAH 6.2\,$\mu$m feature. 
Such a steep flux rise can only be accounted for by PAH emission. Using this fact, we propose a new photometric selection method for galaxies whose mid-IR emission is dominated by PAH emission. This is possible if photometric observations are sufficiently comprehensive enough to trace the PAH  emission feature in the SED, as in the AKARI/IRC all-band photometric survey. We call galaxies selected in this method PAH-selected galaxies. 

In this paper, we utilize the multi-wavelength mid-IR coverage of the NEP-Deep survey with AKARI, in order to 
identify galaxies with strong PAH emission. Sections 2 and 3 describe the data and sample selection, respectively. Our SED-fitting method using the optical to mid-IR wavelength range, and its results are presented in section 4. We discuss the properties of PAH-selected galaxies in section 5 and summarise the results in section 6. Throughout this paper, a flat cosmology with $H_0 = 70$\,km\,s$^{-1}$\,Mpc$^{-1}$ and $\Omega_\Lambda =0.7$ is used. All magnitudes are given in the AB system, unless otherwise stated.

\section{Data}
Here we use the multi-wavelength data set of the AKARI NEP-Deep survey, described in detail in \cite{2008PASJ...60S.517W}. The NEP-Deep survey covers a circular area of 0.38\,deg$^2$, centered at $\alpha = 17^\mathrm{h} 56^\mathrm{m} 01^\mathrm{s}$ and $\delta = 66^\circ 33' 48.4''$ (J2000.0), with all nine IRC bands, i.e.~2.4, 3.2, 4.1, 7.0, 9.0, 11, 15, 18, and 24\,$\mu$m ($N2$, $N3$, $N4$, $S7$, $S9W$, $S11$, $L15$, $L18W$, and $L24$ in conventional band names, respectively). The 5\,$\sigma$ sensitivities in these bands are 9.6$^1$, 7.5$^1$, 5.4{\footnote[1]{Corrected from the values in \cite{2008PASJ...60S.517W}, in which the adopted conversion factors for $N2$, $N3$, and $N4$ bands are not correct. The correct sensitivity is better by a factor of 1.47.}}, 49, 58, 71, 117, 121 and 275\,$\mu$Jy from $N2$ to $L24$, respectively \citep{2008PASJ...60S.517W}. 
Deep optical ($BVRi'z'$ and $NB711$) images with the Subaru/Suprime-cam (S-cam), reaching a limiting magnitude of $B=28$ and $z'=26$ AB mag, are available for a part of the NEP-Deep field with an area of $27'\times 34'$, i.e. one field-of-view of S-cam. We also have ground-based near-IR ($JK_s$) images with KPNO-2.1m/FLAMINGOS for the S-cam field \citep{2007AJ....133.2418I}, reaching a Vega magnitude of $K_s = 19.5$. 
The ground-based near-IR images have higher spatial resolution (FWHM of 1.08$''$) than that of IRC, 
and are therefore quite useful to resolve source confusion in IRC images. 
In this paper, we concentrate on sources in the S-cam field only. 


We performed an optical spectroscopic survey of 242 mid-IR sources with $R\la 24$\,mag in the S-cam field using Keck/DEIMOS (Takagi et al. in prep). Here we use the spectroscopic redshifts for the calibration of photometric redshifts. The optical emission line diagnostics of mid-IR sources will be given elsewhere.

  \begin{figure}
     \resizebox{8cm}{!}{\includegraphics{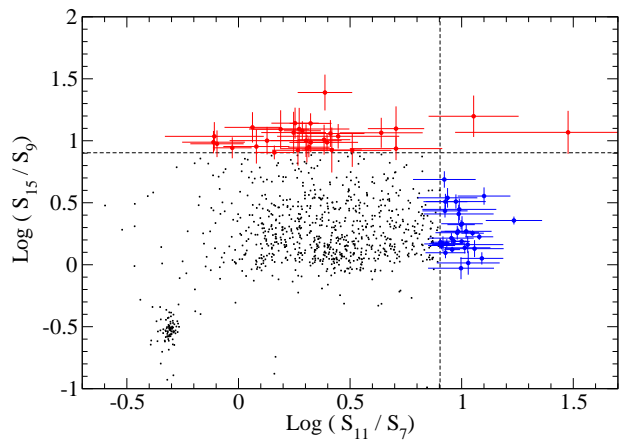}}
 \caption{A colour-colour plot of all-band--detected sources with 15-to-9\,$\mu$m and 11-to-7\,$\mu$m flux ratios. Dashed lines indicate the colour cuts for selecting PAH-selected galaxies. Points with error bars indicate the PAH-selected galaxies. 
}
 \label{colcol}
\end{figure}

  \begin{figure}
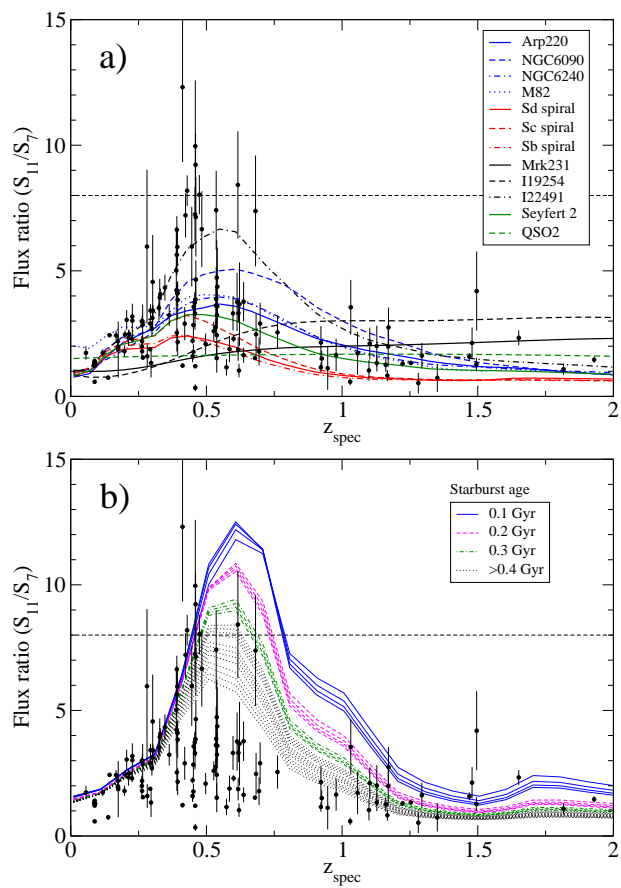

   \resizebox{8cm}{!}{\includegraphics{col11-7_swire.eps}}
      \resizebox{8cm}{!}{\includegraphics{col11-7.eps}}
 \caption{ 11-to-7\,$\mu$m flux ratio as a function of redshift. Solid circles with error bars indicates the spectroscopic sample with $S_7, S_9, S_{11} > 30$\,$\mu$Jy  and $S_{15} > 100$\,$\mu$Jy. a) Flux ratio estimated from the SED template of \cite{2007ApJ...663...81P}. The names of galaxies in the template are given in the legend. b) Same as a), but with the SBURT model with different starburst age and compactness of the starburst region ($\Theta $=1.4, 1.6, 1.8, 2.0) , which controls the optical depth. The MW dust model is adopted.  [{\it See the electronic edition of the Journal for a colour version of this figure}] 
}
 \label{colz_pah1}
\end{figure}

  \begin{figure}
   \resizebox{8cm}{!}{\includegraphics{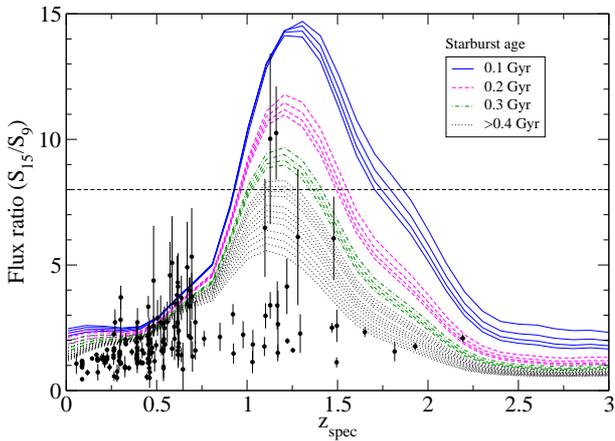}}
 \caption{  Same as Figure \ref{colz_pah1}b), but for 15-to-9\,$\mu$m flux ratio.  }
 \label{colz_pah2}
\end{figure}

\section{Sample selection} 

\subsection{All-band--detected sources}
In order to evaluate and to maximize the new diagnostic capabilities becoming available from the NEP-Deep survey, we have constructed a source catalogue of objects, detected in all nine IRC bands (hereafter all-band--detected sources). We have combined the $S7,S9W,S11$-band images for source detection and used  SExtractor \citep{1996A&AS..117..393B} to make an initial catalogue. 
We then follow \cite{2007PASJ...59S.557T} for photometry with IRC images, in which the aperture photometry with 2 pixel and 3 pixel radii are used for near-IR ($N2,N3,N4$) and mid-IR images ($S7,S9W,S11,L15,L18W,L24$), respectively. Appropriate aperture corrections are then applied. We also follow \cite{2007PASJ...59S.557T} for band merging. Starting from the centroid position of $S11$ sources, we searched for the centroid in the other images. If the angular distance between centroids is less than $3''$, the objects are considered to be the same \citep[see][]{2008PASJ...60S.517W}. Since we have not adopted a specific signal-to-noise ratio for the detection at all of the IRC bands other than $S11$, the resulting catalogue includes less significant detections ($<3\,\sigma$) in some cases. Nevertheless, we include these sources in our all-band--detected sample, since the less significant detections are rare, and  the resulting SED is still well constrained. 

There are 1100 all-band--detected sources in the entire NEP-Deep field, and 630 in the S-cam field. We searched for optical identification using sky positions in the $N2$ band, where the astrometry is based on the Two Micron All Sky Survey (2MASS) and therefore most reliable. The resulting optical identifications were all visually inspected. We found that about 10\,\% of sources have ambiguous optical identifications. Excluding these sources, we obtained 568 all-band--detected sources in the S-cam field with unambiguous optical identification, of which 113 have optical spectra with DEIMOS.

\subsection{PAH-selected galaxies}

At the mid-IR wavelength region, emission from galaxies is usually dominated by small and hot dust grains experiencing the temperature fluctuation. PAHs are believed to be the smallest of those, consisting of $\sim 100$ carbon atoms grouped into  $\sim$10\,$\AA\ $ in dimensions, and having emission bands. PAH emission features have sometimes been used as star formation indicators, which are prominent in starburst galaxies \citep[][]{2006ApJ...653.1129B}, but weak or absent in 
AGNs \citep{2005ApJ...633..706W}. 

Using the very steep flux rise on the blue side of the 6.2\,$\mu$m PAH feature, we attempt to identify 
galaxies whose mid-IR emission is dominated by PAHs, based on AKARI photometric observations. 
The steep flux rise due to PAH emission can be characterized by the flux ratio at the rest-frame wavelengths of approximately 7 to 4\,$\mu$m. For galaxies at $z\sim 0.5$ and 1, we could use 11-to-7\,$\mu$m flux ratio and 15-to-9\,$\mu$m flux ratio, respectively, in order to identify galaxies with prominent PAH emission. For convenience, we hereafter use the term, the PAH-to-continuum flux ratio, for these flux ratios. Figure \ref{colcol} shows a colour-colour plot of all-band--detected sources using these flux ratios. There are galaxies with the PAH-to-continuum flux ratio as high as 10 for both ratios. We hereafter call galaxies with the flux ratio of $>8$ as PAH-selected galaxies. 
 
For star-forming galaxies, it is likely that the flux at the rest-frame 4\,$\mu$m is dominated by the stellar continuum, while PAH 7.7\,$\mu$m feature dominate the flux at the rest-frame 7\,$\mu$m. The PAH luminosity is believed to be a good tracer of total infrared luminosity of star-forming galaxies, and hence the SFR \citep[e.g.][]{1998ApJ...498..579G,1999AJ....118.2625R,2007ApJ...667..149F}. Thus, these flux ratios are roughly proportional to the mass-normalized SFRs, i.e. specific SFR. We hence expect that PAH-selected galaxies are the best candidates of starburst galaxies.
 
In Figure \ref{colz_pah1}a, we show the 11-to-7\,$\mu$m flux ratio as a function of redshift, using the SED template of various galaxy types -- the SWIRE template library  \citep{2007ApJ...663...81P}. This template confirms that the 11-to-7\,$\mu$m flux ratio has a peak at $z\sim 0.5$, although no SEDs in the template can reproduce the flux ratios greater than 8. We note that the SWIRE template library has only a few SEDs of starburst-dominated galaxies. In order to explain the high flux ratio of PAH-selected galaxies, we may need more comprehensive SED templates of starburst galaxies. 

For such SED template of starburst galaxies, we adopt an evolutionary starburst SED model with radiative transfer 
\citep[][SBURT]{2003MNRAS.340..813T}. This is an SED model for starburst regions, assumed to be spherical and have a concentrated stellar distribution and homogeneous dust distribution. SBURT adopts a stellar population synthesis model of \cite{1997A&A...320...41K} as input. The amount of dust is calculated, based on the chemical evolution assuming gas infall. The time-scale of gas infall and of star formation is assumed to be 0.1\,Gyr~\footnote{The model SED of starbursts scales with the time-scale of gas infall and star formation  \citep{2003MNRAS.340..813T}. Therefore, the starburst age from the SBURT model fitting should be considered only in the relative sense.}. Thus, the SBURT model can be regarded as a population synthesis model with physically consistent treatment of dust extinction and re-emission. The optical depth of the system is controlled by the compactness ($\Theta$) of the starburst region. In the SBURT model, there are three choices of the extinction curve, corresponding to the dust model for the Milky Way (MW), Large or Small Magellanic Clouds (LMC or SMC). The fraction of PAHs in dust is 5, 1 and 0.5\,\% in the MW, LMC, and SMC dust model, respectively. Thus, models with the MW dust has the largest PAH-to-continuum flux ratio. See \cite{2003MNRAS.340..813T,2003PASJ...55..385T} for more details of SBURT. 

In Figure \ref{colz_pah1}b and \ref{colz_pah2}, we show the 11-to-7\,$\mu$m and 15-to-9\,$\mu$m flux ratios estimated from the SBURT model, respectively. We found that the MW dust model is required, in order to reproduce high flux ratios of PAH-selected galaxies. According to the model, younger galaxies may have larger flux ratios. This is because the contribution from the stellar component to the $4\,\mu$m continuum is less for younger starbursts. The flux ratios of greater than 8 are satisfied with the model younger than 0.4\,Gyr. However, this is only one possible interpretation, since the dust model adopted has only three choices (MW, LMC or SMC), and limited capability of reproducing large variations of mid-IR properties. Specifically, this model does not take the ionization status of PAHs into account, which significantly affects the optical constants of PAHs \citep[e.g.][]{2001ApJ...554..778L}.

Among all-band--detected sources, we identify 56 (39) PAH-selected galaxies from 11-to-7\,$\mu$m (15-to-9\,$\mu$m) flux ratio in the entire NEP-Deep field, of which 38 (18) lie in the S-cam field with unambiguous optical identification. We include sources with less significant detection at the rest-frame 4\,$\mu$m, if the 2.5\,$\sigma$ upper limit is consistent with the selection criterion. In Table A.1, we tabulate the IRC photometry of PAH-selected galaxies.

\section{Photometric redshifts and SED fits}
\subsection{Method}

We fit the optical-to-MIR SEDs of all-band--detected sources with the SBURT model. 
We use the model with the starburst age of 0.01 -- 0.6\,Gyr, which is enough to represent the starburst phase. 
The SBURT model originally covers the wavelength range from UV to submillimetre. \cite{2007MNRAS.381.1154T} empirically added the radio component to reproduce the observed radio-IR correlation. This model does not include any emission from AGN. Thus, it is expected that the SBURT model underpredicts the mid-IR fluxes from AGN hosts, which usually have hot dust excess compared to pure starburst galaxies \citep{1997ApJS..108..229H}.

We adopt the same SED fitting method as in \cite{2007PASJ...59S.557T}, using a standard $\chi^2$ minimization technique. \cite{2007PASJ...59S.557T} adopted relatively large minimum photometric errors, 20 -- 30\,\% depending on the band. Since the photometric calibration of IRC is recently improved \citep{2008PASJ...60S.375T}, we here set the minimum photometric errors to 15\,\% for all IRC bands. As in \cite{2007PASJ...59S.557T}, we quadratically added an additional 20\,\% error for data at the rest-frame UV wavelengths, $<4000$\,\AA, in order to account for the uncertainty of the extinction curve. 

We apply the SBURT model for all-band--detected sources. By nature, all-band--detected sources are heterogeneous
sample, even including stars. Since we use a starburst SED template, SBURT, for this sample, it is important to discriminate starburst candidates from other type of sources. To do this, we use the goodness of SBURT fitting and reject sources with no acceptable SBURT model in further analyses. We reject the best-fitting SBURT model if the value of resulting $\chi^2$ is so large that the corresponding probability is less than $1$\,\%.

  \begin{figure}
  \resizebox{8cm}{!}{\includegraphics{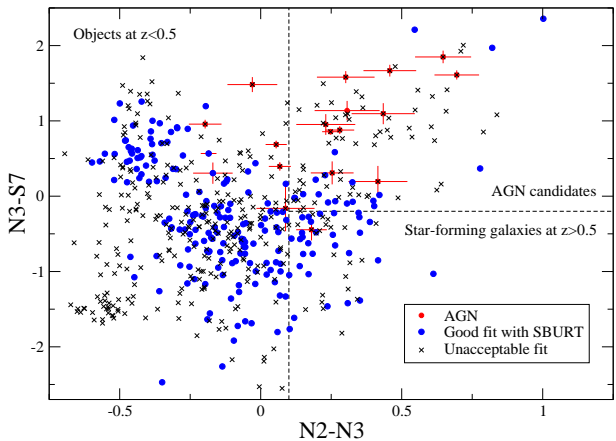}}
 \caption{Colour-colour plot with $N3-S7$ versus $N2-N3$. Solid circles and small crosses indicate the good 
 and bad-fit sample, respectively. Solid circles with error bars represent optically identified AGN. 
}
 \label{colcol2}
\end{figure}

  \begin{figure}
  \resizebox{8cm}{!}{\includegraphics{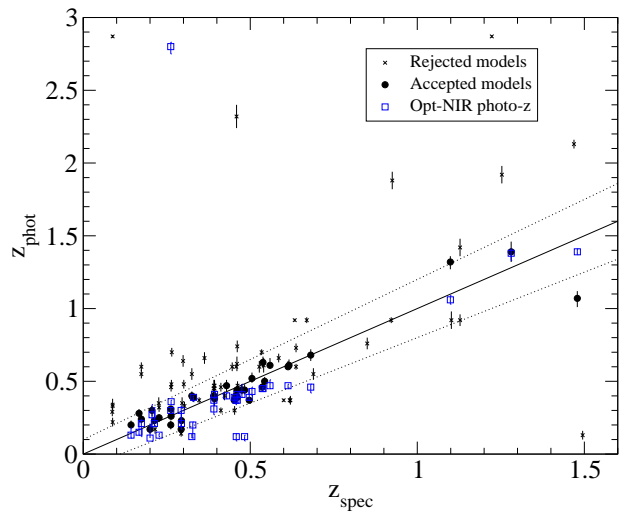}}
 \caption{Photometric versus spectroscopic redshifts. Solid circles indicate the redshifts of the good-fit sample, while small crosses are for the bad-fit sample. Dotted lines represent $\frac{\Delta z}{(1+z)} = \pm 0.1$. Squares indicate the photometric redshifts of the good-fit sample, based on the ground-based optical-NIR photometry only. 
}
 \label{photoz}
\end{figure}

  \begin{figure}
  \resizebox{8cm}{!}{\includegraphics{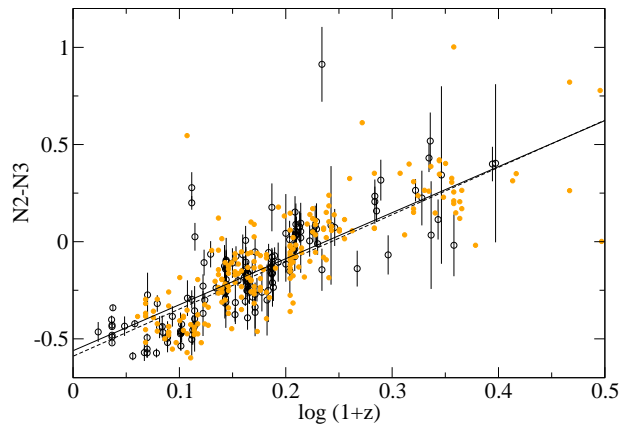}}
 \caption{Correlation between $N2-N3$ colour and redshift for both photometric and spectroscopic samples. 
 Open circles with error bars indicate the spectroscopic sample. Solid circles represent the photometric sample. 
 Solid and dashed lines are the regression line for photometric and spectroscopic sample, respectively. 
}
 \label{colz_corr}
\end{figure}

\subsection{Fitting results}
\subsubsection{All-band--detected sources}
 
We performed SED-fitting for 568 all-band--detected sources using the SBURT model. 
We obtained good fits for $\sim 40$\,\% (211/568) of all-band--detected sources. On the other hand, using the SWIRE template library, instead of SBURT, we obtained good fits for only $\sim$10\,\% (69/568) of sources, although it covers a wide range of SED type. Note that the SWIRE library has only a few SEDs of starbursts. We think that a large variety of starburst SEDs, as in the SBURT model, is necessary to reproduce a good fraction of observed UV--mid-IR SEDs of IR-selected galaxies. 

In order to show what kind of sources are rejected in the SBURT fitting, we plot both accepted and rejected samples in a colour-colour plot of $N3-S7$ vs $N2-N3$ in Figure \ref{colcol2}. \cite{2007PASJ...59S.557T} use this colour-colour plot to discriminate AGNs from normal star-forming galaxies. AGNs identified with optical spectroscopy have red colours in both $N2-N3$ and $N3-S7$. The accepted fits avoid this AGN colour region, because the SBURT model does not have any AGN component. The accepted sample also avoids the clump of objects with blue colours at $N2-N3\simeq 0.6$ and $N3-S7\simeq -1.5$, which are stars. The rest of the rejected sample have similar $N2-N3$ and $N3-S7$ colours to those of the accepted sample. These include quiescent spiral galaxies, which have systematically redder optical--near-IR colours than any of the SBURT models. As we discuss in Section 5.1, limitations on the dust model may also cause rejection of the best-fitting model.

In order to derive any firm conclusions, photometric redshifts of the good-fit sample should be accurate enough. 
Figure \ref{photoz} shows the comparison of photometric redshifts to spectroscopic redshifts. Solid circles and crosses represent galaxies with acceptable and rejected fits, respectively. Defining a catastrophic error by $\frac{\Delta z}{(1+z)} > 0.2$, we find no catastrophic errors and $\sigma_\frac{\Delta z}{(1+z)} = 0.034$ for the good-fit sample, and 21/68 catastrophic errors for the bad-fit sample. Thus, the goodness of the SBURT fit is actually a good measure of the accuracy of photometric redshifts. 

In Figure \ref{photoz}, we also plot photometric redshifts derived from ground-based optical-NIR ($BVRi'z'JK$) bands only, using a widely-used photometric redshift code, Hyperz \citep{2000A&A...363..476B}. In Hyperz, we adopt a population synthesis model of \cite{2003MNRAS.344.1000B}, which is distributed along with the code, and the extinction curve of starbursts \citep{2000ApJ...533..682C} for dust reddening. In order to compare the photometric redshifts with AKARI bands to those without AKARI, we show the good-fit sample (with SBURT fitting) only in Figure \ref{photoz}. We find one catastrophic error in ground-based photometric redshifts, while this galaxy has a reasonable photometric redshift with AKARI bands. Furthermore, ground-based photometric redshifts seem to have systematic errors; i.e. at $z\sim 0.5$, most of photometric redshifts are lower than the spectroscopic redshifts. Such errors can be removed by using AKARI bands.

In order to check the statistical reliability of our photometric redshifts, we use a correlation between the NIR colour $N2-N3$ and redshift \citep{2007PASJ...59S.557T}. 
With this test, we could detect systematic effects, if any, due to the usage of mid-IR fluxes for deriving photometric redshifts. 
In Figure \ref{colz_corr}, we show the correlations for both spectroscopic and photometric samples. For the photometric sample, we obtain the correlation of $(N2-N3) = (2.37\pm0.12) \log(1+z) - 0.56$, which is consistent with that of the spectroscopic sample, i.e. $(N2-N3) = (2.43\pm0.05) \log(1+z) - 0.59$. Therefore, we find no systematic errors in our photometric redshifts.

\subsubsection{PAH-selected galaxies}

We here show more details of SED fitting results specifically for PAH-selected galaxies. In Figure \ref{sedz05good} -- \ref{sedz1bad}, we show the best-fitting (both accepted and rejected) SED models for PAH-selected galaxies at $z\sim 0.5$ and 1. We obtained acceptable SED fits for 22/38 and 10/18 of the PAH-selected galaxies at $z\sim 0.5$ and 1, respectively. Despite the fact that PAH-selected galaxies are starbursts, the SBURT model does not provide acceptable fits for 40\,\% of PAH-selected galaxies, although the success rate of the fitting is better than that of whole sample. This topic is discussed in detail in the next section.  

\begin{figure}
     \resizebox{8.5cm}{!}{\includegraphics{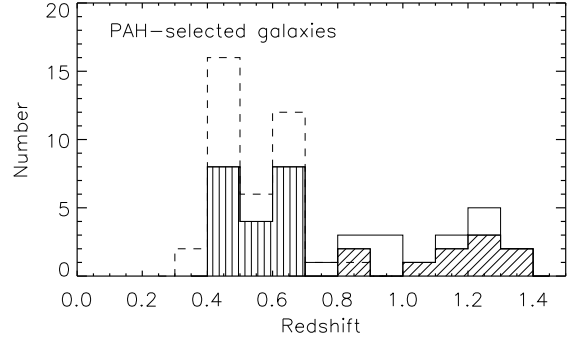}}
 \caption{Redshift distribution of PAH-selected galaxies. Vertically and diagonally shaded histograms represent 11-to-7\,$\mu$m and 15-to-9\,$\mu$m selected galaxies with good SED fits, respectively. Histograms with thin dashed and solid lines indicate the redshift distribution including the bad-fit sample.  Spectroscopic redshifis are adopted if available. 
}
 \label{zhist}
\end{figure}

We show the redshift distributions of PAH-selected galaxies in Figure \ref{zhist}.  They are consistent with the expectation from the model templates, shown in Figure \ref{colz_pah1} and \ref{colz_pah2}. There are no systematic differences in photometric redshifts between the good- and bad-fit samples. This may indicate that the resulting large $\chi^2$ is not due to an incorrect redshift. The total-IR luminosity from the SED model is indicated in each panel of Figure \ref{sedz05good} -- \ref{sedz1bad}. We note that PAH-selected galaxies at $z\sim 1$ include ULIRGs. This type of ULIRGs can be found only at high redshifts (see Section \ref{lumRel}).

  \begin{figure}
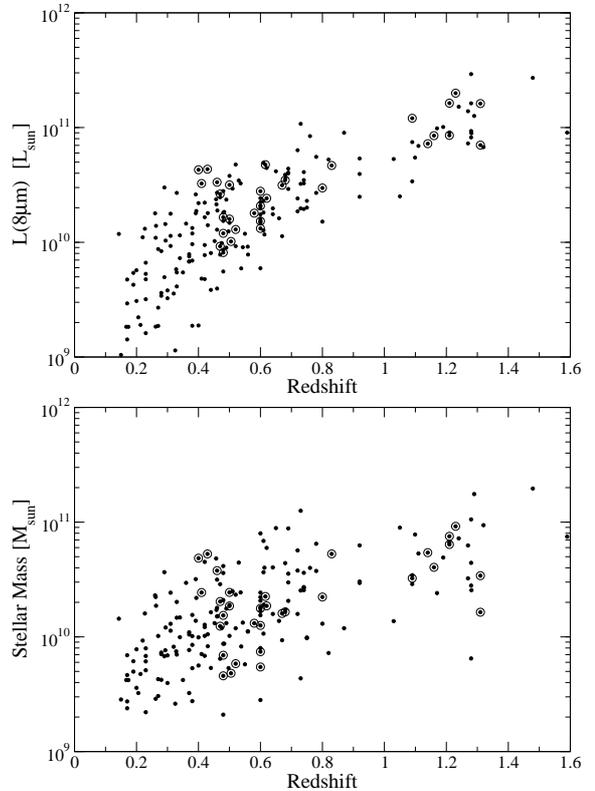

      \resizebox{7.5cm}{!}{\includegraphics{z_l8um.eps}}
  \resizebox{7.5cm}{!}{\includegraphics{z_mstar.eps}}
 \caption{
8\,$\mu$m luminosity and stellar mass as a function of redshift. 
All-band--detected sources and PAH-selected galaxies are represented by small dots and open circles, respectively. 
We show galaxies with good SED fits only. 
}
 \label{mass}
\end{figure}


\section{Discussion}

\subsection{Nature of PAH-selected galaxies}

  \begin{figure}
  \resizebox{8.5cm}{!}{\includegraphics{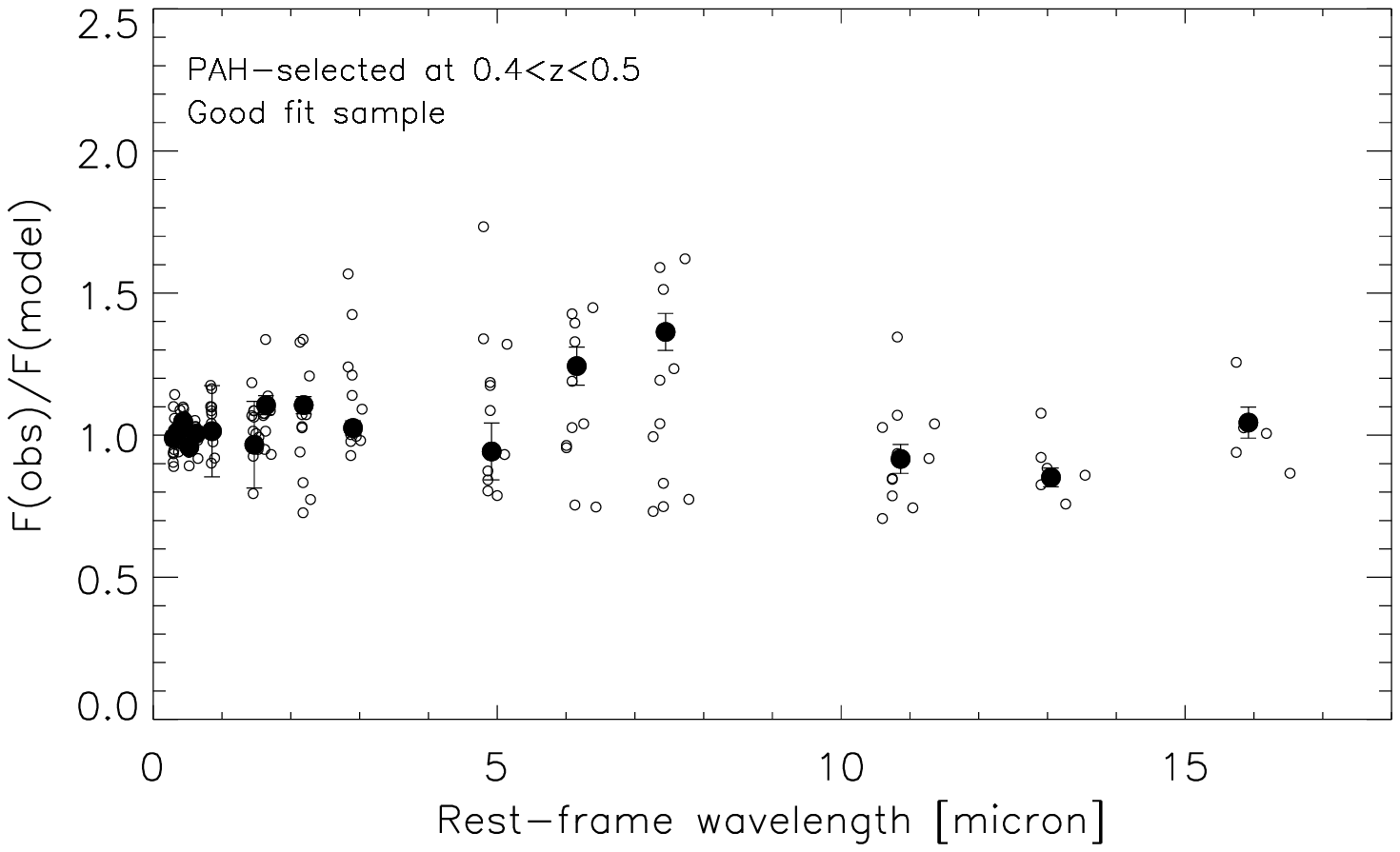}}
    \resizebox{8.5cm}{!}{\includegraphics{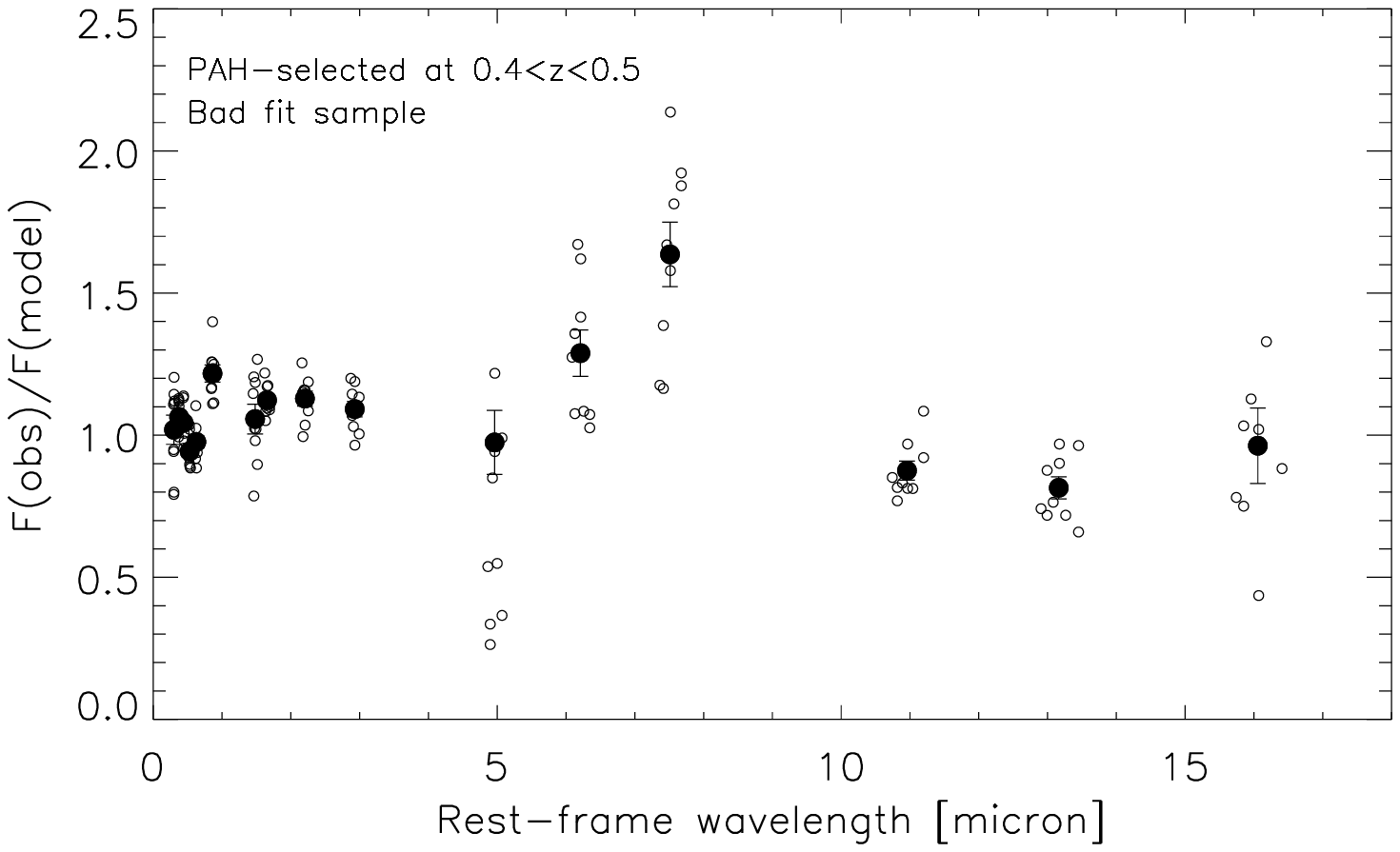}}
 \caption{
 Ratio of observed to model fluxes as a function of rest-frame wavelengths for PAH-selected 
 galaxies at $0.4<z<0.5$.  Open and solid circles indicate individual galaxies and the weighted average for each photometric band, assuming the mean redshift of 0.46 and 0.45 for good and bad fit sample, respectively. 
}
 \label{pah_excess05}
\end{figure}

In Figure \ref{mass}, we show 8\,$\mu$m luminosities and stellar masses of PAH-selected galaxies, along with those of all-band--detected sources. The 8\,$\mu$m luminosity is derived from the observed flux in the IRC bands closest to the reset-frame 8\,$\mu$m and the luminosity distance from photometric redshifts. Some of PAH-selected galaxies at $z\sim 1$ have $L_{8\,\mu m}$$ > 10^{11}$L$_\odot$, which correspond to the luminosity of ULIRGs  
for a typical starburst SED \citep{2008ApJ...686..127W}. 

Stellar masses are derived from the best-fitting SED model when it is accepted. The adopted initial mass function (IMF) is a top-heavy one with the power law index of $x=1.10$ (the Salpeter IMF has $x=1.35$), and the lower and upper mass limits are 0.1 and 60\,$M_\odot$ \citep[see also][]{2004MNRAS.355..424T}. For the Salpeter IMF, stellar mass increases by a factor of 2.  
Stellar masses thus derived should be regarded as a lower limit, since the contribution from evolved stellar populations is not included. 
Most of PAH-selected galaxies at $z\sim 1$ have stellar masses of $>3\times 10^{10}$\,M$_\odot$. Even some fraction of PAH-selected galaxies at $z\sim 0.5$ have similar stellar masses. 
What are descendants of such massive PAH-selected, i.e. starburst galaxies? At redshift $z\sim 2$, massive starburst galaxies have been found mainly as submillimetre galaxies \citep[SMGs; e.g.][]{2005ApJ...635..853B} and they are good candidates of massive spheroids \citep[][]{2004ApJ...611..725B,2004ApJ...616...71S,2007ApJ...671..303B,2008ApJ...680..246T}. However, PAH-selected galaxies are not as massive as SMGs. Their luminosity is typically an order of magnitude less than that of SMGs, corresponding to luminous infrared galaxies (LIRGs\footnote{Galaxies with a total infrared (8 -- 1000\,$\mu$m) luminosity of 10$^{11\mathrm{-}12}$\,L\,$_\odot$.}). Recent high resolution images of LIRGs at $z\sim 1$ show that the majority of LIRGs have disk morphology \citep{2007A&A...468...33E,2008AJ....135.1207M}, although the sample size is still small. Clearly, we need more comprehensive study to identify what triggers starburst activity of PAH-selected galaxies, in order to constrain their evolutionary path. 


What is the main cause of a large PAH-to-continuum flux ratio? As shown in Figure \ref{colz_pah1} and \ref{colz_pah2}, the SBURT model predicts that galaxies with large PAH-to-continuum flux ratios are young starbursts with the starburst age of 
$\la 0.4$\,Gyr. For the whole sample of PAH-selected galaxies with good SED fits, we derive the mean starburst age of 0.4\,Gyr. We note that, in the SED fitting, the starburst age depends not only on the PAH-to-continuum flux ratio, but also on the optical colours \citep{1999ApJ...523..107T}. The lack of very young starbursts in the sample may be accounted for by a selection effect, i.e. such young systems are hard to be observed due to a short life time. 
 
Indeed, star-forming regions, i.e. young stellar systems in a galaxy have a large PAH-to-continuum flux ratio, as shown in Spitzer/IRAC images of nearby galaxies in the literature. \cite{2004ApJS..154..193W} study the Antennae galaxies (NGC4038/4039) and show the map of the 8.0\footnote{Non-stellar emission, i.e. $\sim5$\,\% of stellar emission is subtracted from the total flux}-to-4.5\,$\mu$m flux ratio, which is close to the PAH-to-continuum flux ratio in our definition. This flux ratio is quite high only at the overlap region of the two disks where the most of IR emission come from. They estimate that such active star-forming region comprise only $\sim$10\,\% of the total stellar mass of the entire system; i.e. the activity is localized rather than global. On the other hand, PAH-selected galaxies should have global star-forming activity, since the flux ratio for the entire system is high. 

For fair discussion, we should also pay attention to the case with {\it no} acceptable fits and clarify the cause of poor fits. 
In Figure \ref{pah_excess05}, we show the observed-to-model flux ratio as a function of rest-frame wavelength for PAH-selected galaxies at $0.4<z<0.5$. All of the rejected models, but one in Figure \ref{pah_excess05} have a starburst age of 0.6\,Gyr, i.e. the oldest and reddest model in the template. This means that the model-fitting fails for galaxies with relatively red optical colours, but with strong PAH emission (see also Ohyama et al. in preparation). Since the SBURT model is a rather simple model of starbursts, which have a single starburst region and starburst stellar population only, it is not surprising to find its limitation. 

Figure \ref{pah_excess05} shows that a large discrepancy between observations and the model can be found at the 5 -- 7\,$\mu$m range with sharp rising trend towards longer wavelength. We can see the same trend for the good-fit sample as well, although it is less significant. This indicates that the dust model adopted in the SBURT may need to be modified. One possibility is the ionization of PAHs. The optical constant of PAHs in the SBURT model is for neutral PAHs, which may not be true for every PAHs. \cite{2001ApJ...554..778L} suggest that ionized PAHs have systematically higher absorption cross section at 4 -- 10\,$\mu$m, compared to neutral PAHs. If some fraction of PAHs are ionized, the discrepancy found in the SED fitting may be recovered.

Another possibility is that stars dominating optical light is not the same as those responsible for strong PAH emission. As starbursts age, dust heating would be mostly dominated by newly formed stars in molecular clouds, while most of optical light come from stars having already exited from molecular clouds; 
 i.e. older stars are less attenuated and dominate optical light, while younger stars embedded in dense molecular clouds dominate infrared light.  
Such age-dependent dust attenuation would be more effective in the later stage of starbursts. This may explain why we find larger discrepancy in optically red, probably old, starbursts.

Above two possibilities could be discriminated with the far-IR photometry. The former, the case of ionized PAHs, would change only mid-IR luminosity and does not affect the far-IR luminosity. If the latter (age-dependent attenuation) is the case, the current model would underestimate not only mid-IR luminosity but also far-IR luminosity. Thus, expected PAH-to-total-IR luminosity depends on the scenario. Sensitive far-IR photometry of red PAH-selected galaxies, e.g. with the Herschel Space Observatory, could have diagnostic power to identify the main cause of the discrepancy.



  \begin{figure*}
  \resizebox{17cm}{!}{\includegraphics{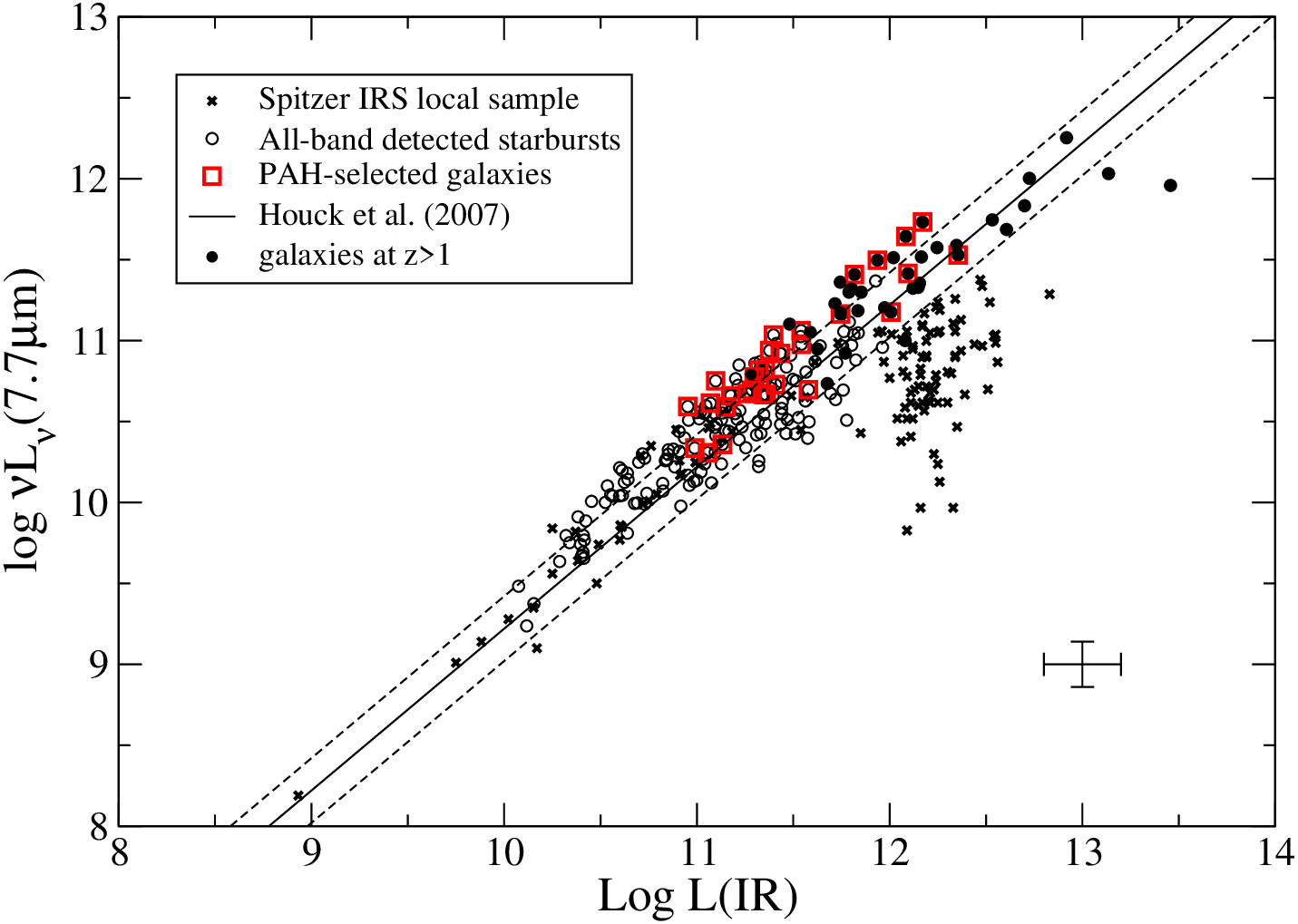}}
 \caption{7.7\,$\mu$m peak luminosity versus infrared luminosity. Solid and dashed lines represent the correlation between these luminosities found for less luminous starburst galaxies and its 1\,$\sigma$ uncertainty \citep{2007ApJ...671..323H}. Legend describes symbols for  local galaxies from Spitzer IRS sample from \cite{2008ApJ...686..127W}, all-band--detected sources, and PAH-selected galaxies at both $z\sim 0.5$ and 1.  Solid circles indicate galaxies with $z>1$. Typical uncertainty of both luminosities is indicated at the bottom right corner. 
}
 \label{Lir_pah}
\end{figure*}

\subsection{PAH-to-total IR luminosity relation} \label{lumRel}

The PAH luminosity of local ULIRGs is less than that expected from the correlation of PAH and total IR luminosity for less luminous starburst galaxies. In Figure \ref{Lir_pah}, we show this luminosity relation for galaxies at $z<0.2$ with crosses. The PAH luminosity is represented with the PAH 7.7\,$\mu$m peak luminosity taken from \cite{2008ApJ...686..127W}. The total infrared luminosity is taken from the references in \cite{2008ApJ...686..127W}, and converted to the cosmology adopted here. The 7.7\,$\mu$m peak luminosity of ULIRGs is systematically below the correlation for local starbursts. A most plausible explanation of this systematic effect is an AGN contribution to the total IR luminosity. \cite{2007ApJS..171...72I} claim that 30-50\,\% of optical non-Seyfert ULIRGs at $z<0.15$ have dust-enshrouded AGN. Combined with optical Seyfert ULIRGs, their results indicate that $\sim 50$\,\% of ULIRGs at $z<0.15$ harbor AGN \citep[see also][]{1998ApJ...498..579G,1998ApJ...505L.103L}.  However, Figure \ref{Lir_pah} shows that none of $z<0.2$ ULIRGs are consistent with the pure starburst case. This may indicate another explanation for the lower 7.7\,$\mu$m luminosity of ULIRGs. If the starburst region is heavily obscured, the absorption of PAH emissions by silicate dust may not be negligible. 

High redshift ULIRGs may behave differently, owing to possible evolutionary effects. We investigate the PAH-to-total IR luminosity relation of the all-band--detected sources, including PAH-selected galaxies. The 7.7\,$\mu$m peak luminosities are estimated from the IRC photometry by the following method. For a local starburst spectral template of \cite{2006ApJ...653.1129B}, filter-convolved flux at the rest-frame 8\,$\mu$m is approximately half of the  7.7\,$\mu$m peak flux; 0.50 in the $L15$ filter at $z=1$ and 0.42 for the  $S11$ filter at $z=0.5$ for example. Here we adopt 0.50 for the conversion factor, since $z\sim 1$ sources are our prime targets. This estimate would be accurate within a  30\,\% uncertainty, as far as the 8\,$\mu$m emission is dominated by starbursts.  
We note that intrinsic spectral variation at 5 -- 8\,$\mu$m is found to be small in starbursts \citep{2008MNRAS.385L.130N}. 
The rest-frame 8\,$\mu$m luminosities are estimated from the observed fluxes at the band closest to the rest-frame 8\,$\mu$m and the luminosity distance derived from the redshift estimates. We have not done any interpolation or $k$-correction in this step, considering the uncertainty of the photometric redshift. The total IR luminosities are derived from the best-fitting SED model. Therefore, we consider the good-fit sample only.  
A well-known tight correlation between the global far-IR and radio emission allows us to check the reliability of estimated total-IR luminosity from the SED fitting. Among all-band--detected sources, we detected five galaxies in our 1.4\,GHz observation with WSRT  (White et al. 2010 in prep), which covers a part of the NEP-Deep field. Comparing the observed radio fluxes to those predicted from the best-fitting SBURT models based on the far-IR/radio correlation, we estimate that the derived total-IR luminosities have uncertainties of 50\,\%. 
Since the derivations of both the 7.7\,$\mu$m and total IR luminosity assume that starbursts dominate the infrared emission, the results for PAH-selected galaxies, the most plausible starburst candidates in our sample, should be most reliable.

As shown in Figure \ref{Lir_pah}, some PAH-selected galaxies at $z\sim 1$ are ULIRGs. Their 7.7\,$\mu$m
peak luminosities are larger than local ULIRGs with a similar luminosity, and consistent with the PAH-to-total IR luminosity ratio of less luminous starbursts. They have no local counterparts and we specifically call 
these PAH-selected ULIRGs ``PAH-luminous galaxies''. Total IR luminosity of individual galaxies can be found in the legend of Figure \ref{sedz05good} -- \ref{sedz1bad}. In our sample, we find no PAH-luminous galaxies at $z\sim 0.5$.

Normal starbursts, which have unabsorbed PAH emission features and no AGN contribution, can reach higher luminosity at higher redshifts. \cite{2008ApJ...686..127W} reach the same conclusion by compiling various Spitzer IRS spectroscopic samples. \cite{2008ApJ...675..262R} investigate the rest-frame 8\,$\mu$m luminosity of galaxies at $z\sim 2$ as a function of total IR luminosity, and find a similar trend. Using Spitzer/IRS, \cite{2009ApJ...700..183H} also report ULIRGs with high PAH-to-total IR luminosity ratio at $z\sim 2$. Our results show that  there are extreme starbursts with ULIRG luminosity even at $z\sim 1$. 

  \begin{figure}
  \resizebox{8.5cm}{!}{\includegraphics{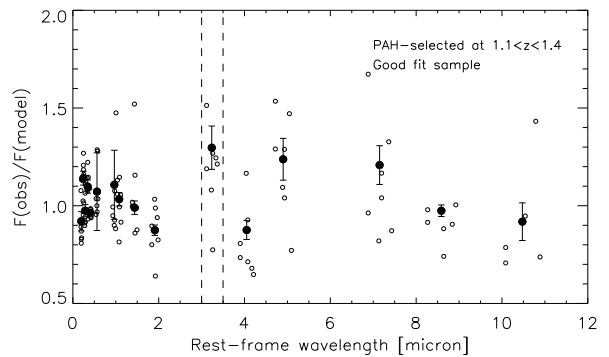}}
 \caption{Same as Figure \ref{pah_excess05} but for PAH-selected galaxies at $1.1<z<1.4$. 
 Open and solid circles indicate individual galaxies and the average for each photometric band, assuming a
 mean redshift of 1.22. Dashed lines bracket the 3.3\,$\mu$m region. 
}
 \label{pah_excess}
\end{figure}

\subsection{Photometric detection of PAH 3.3\,$\mu$m feature?}

For galaxies at $z\sim 1$, the PAH 3.3\,$\mu$m feature falls into $S7$ band. PAH-selected galaxies 
at $z\sim 1$ are good candidates for detecting this feature, because of both redshift and PAH prominence. 
\cite{2008ApJ...681..258M} made a similar attempt using Spitzer IRAC photometry for galaxies at $0.6<z<0.9$, 
and detected an excess over the stellar continuum, which may originate from the PAH 3.3\,$\mu$m feature. 

In Figure \ref{sedz1good}, we can see that some objects, i.e. ID2711, 2836, 2932, 3033, 4956, 
have blue $S7-S9W$ colours, which might be caused by PAH 3.3\,$\mu$m feature. 
Surprisingly, this excess is notable even if compared with the model flux which already 
includes the contribution from 3.3\,$\mu$m feature. We show the comparison of observed fluxes to model 
fluxes in Figure \ref{pah_excess}. Here we use the sample of PAH-selected galaxies at $1.1<z<1.4$ with 
good SED fits. The average observed-to-model flux ratios range from 0.75 to 1.23, depending on the wavelength. 
At PAH-dominated wavelengths, i.e. $>7$\,$\mu$m, the scatter is large and the ratios are marginally consistent with the model. 
At the rest-frame 3\,$\mu$m band, including PAH 3.3\,$\mu$m feature, the flux ratio is systematically high by a factor of 1.23 on average. If this excess is accounted for by a systematically strong 3.3\,$\mu$m feature only, this feature must 
be unusually strong by a factor of 3 -- 7, compared with the model.
In the literature, no galaxies with such strong 3.3\,$\mu$m feature can be found to our 
knowledge. 
This anomaly could originate from the continuum emission, rather than PAH 3.3\,$\mu$m feature. 
However, we stress that, for galaxies at $z\sim 0.5$, the SBURT model reproduces the rest-frame 1 -- 3\,$\mu$m continuum emission well, as shown in Figure \ref{pah_excess05}.

It may be possible that the model lacks additional very hot dust component which contribute to the rest frame 3--4\,$\mu$m emission. Such a near-IR excess with a colour temperature of $\sim$10$^3$\,K is reported in study of nearby galaxies \citep[e.g.][]{2003ApJ...588..199L,2006A&A...453..969F}. However, we note that galaxies with large 3\,$\mu$m flux tend to have small 4\,$\mu$m flux as well, compared with the model. If the additional component has a gray body emission, it is difficult to explain such a trend. Furthermore, PAH-selected galaxies at $z\sim 0.5$ show no systematic near-IR excess, compared to the model, as shown in Figure \ref{pah_excess05}. 

We need more careful investigation of the origin of the flux anomaly at this wavelength range. In fact, the model does not take into account the absorption features due to Hydrogenated Amorphous Carbon (HAC) and/or H$_2$O. Unfortunately, we have to wait for the launch of JWST \citep{2008AdSpR..41.1983C} and SPICA \citep{2004AdSpR..34..645N} to obtain mid-IR spectra of these galaxies and reveal the origin of this flux anomaly.

\section{Summary} 

Using a multi-wavelength mid-IR survey, the AKARI NEP-Deep, we have constructed a catalogue of all-band--detected sources. From this catalogue, we photometrically identified galaxies whose mid-IR emission is dominated by PAH emission. These PAH-selected galaxies have large PAH-to-continuum flux ratio, i.e. 11-to-7\,$\mu$m and 15-to-9\,$\mu$m flux ratios at $z\sim 0.5$ and 1, respectively. PAH-selected galaxies are the best candidates of starburst galaxies at these redshifts. Some of PAH-selected galaxies have stellar masses of $>3\times 10^{10}$\,M$_\odot$, and could be progenitors of present-day massive galaxies, which are mostly spheroids.  

An evolutionary SED model of starbursts, SBURT, reproduces observed optical-to-mid-IR SEDs of more than half of PAH-selected galaxies. According to the SED model, the PAH-to-continuum flux ratio of $>8$ can be explained with the starburst age of 0.4\,Gyr or younger. The average starburst age from the SED fitting is 0.4\,Gyr, which is marginally consistent with the PAH-to-continuum flux ratio. The lack of young starbursts may be the selection effect due to a short life time of young starbursts. On the other hand, SBURT has a difficulty to reproduce large PAH fluxes of optically red PAH-selected galaxies, which are probably evolved starbursts. This may require the SED model to include the age-dependent extinction, and/or to adopt improved optical properties of PAHs. 

At $z\sim 1$, the infrared luminosity of some PAH-selected galaxies corresponds to that of ULIRGs, and we call them PAH-luminous galaxies. They have large PAH luminosity, compared with local ULIRGs, but the PAH-to-total IR luminosity ratio is comparable to that of  less luminous starbursts.  
PAH-luminous galaxies seem to be a unique galaxy population at high redshifts. 
The number density of PAH-luminous galaxies would be given elsewhere, using whole area of the NEP-Deep survey. 

There is a hint that the PAH 3.3\,$\mu$m feature is quite strong in PAH-luminous galaxies by a factor of 3 -- 7, but this needs to be confirmed with next generation infrared telescopes, such as JWST and SPICA, with infrared spectroscopy.

\section*{Acknowledgements} 
We would like to thank all the AKARI team members for their extensive 
efforts. We also appreciate careful reading and constructive comments of the referee. 
This work is supported by the Japan Society for the Promotion 
of Science (JSPS; grant number 18$\cdot$7747). 
This research is based on observations with AKARI, a JAXA project with the participation of ESA.
This research is partly supported with the Grant-in-Aid for 
Scientific Reserch (21340042) from the JSPS.
\bibliography{reference}

\begin{appendix}

\section{SED fitting results}
\begin{figure*}
 \resizebox{19cm}{!}{\includegraphics{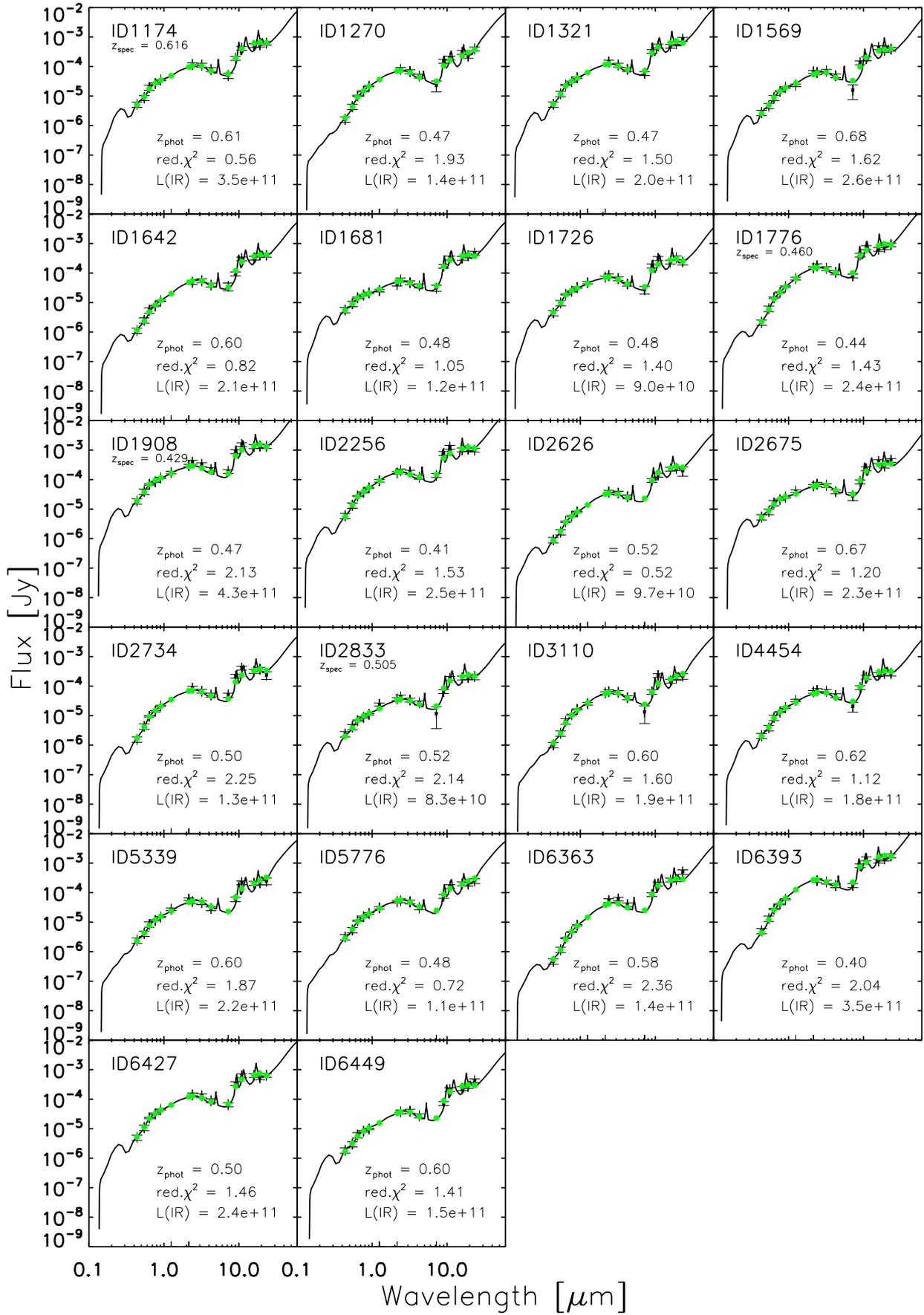}}
 \caption{Results of SED fitting for PAH-selected galaxies at $z\sim 0.5$. We here show the good-fit sample only. 
 Data with error bars indicate the observed flux and errors adopted for the SED fitting. Solid circles are filter 
 convolved fluxes of the best fit model. Spectroscopic redshifts are reported if available. 
}
 \label{sedz05good}
\end{figure*}

\begin{figure*}
  \resizebox{19cm}{!}{\includegraphics{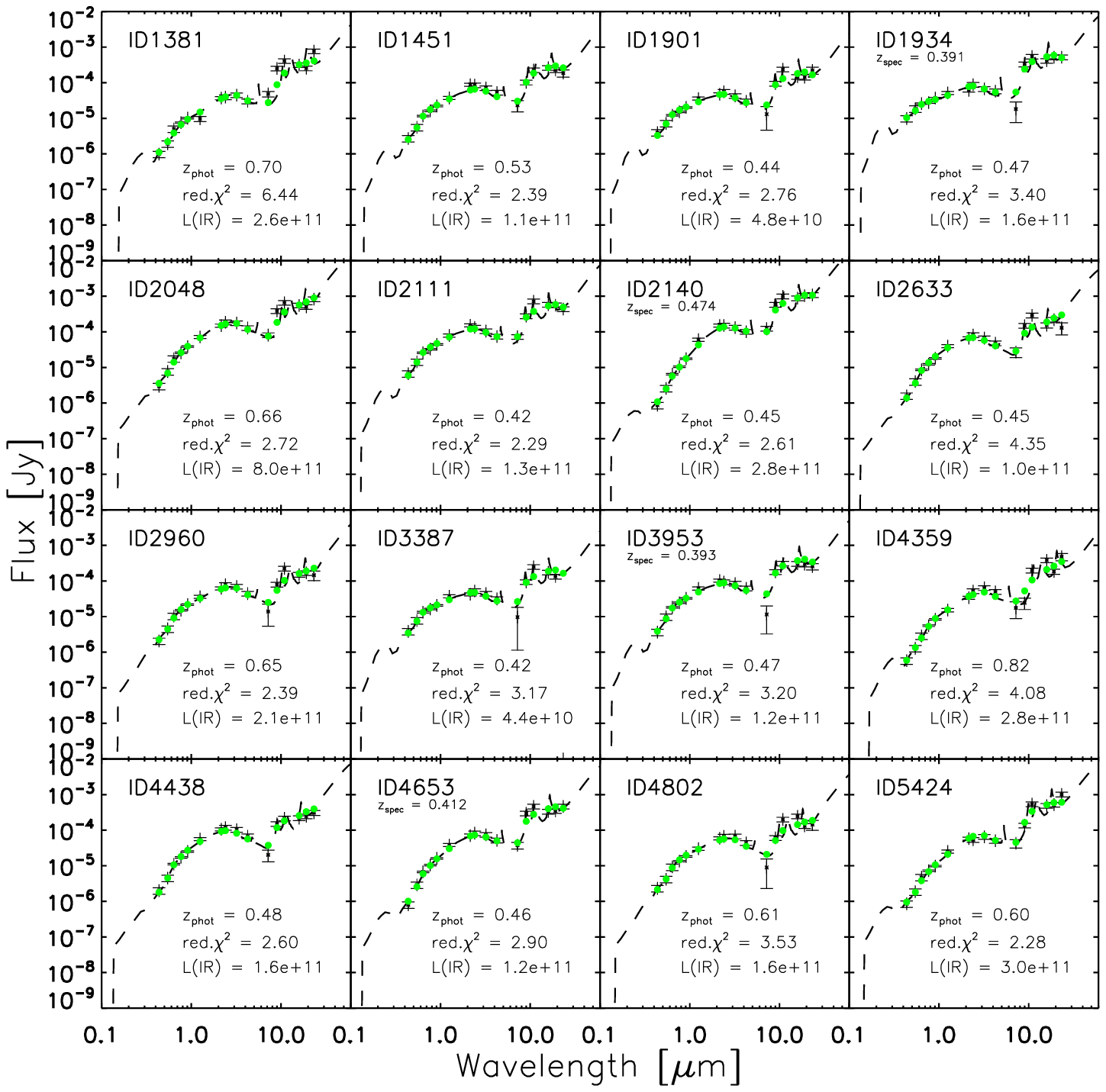}}
 \caption{Same as Figure \ref{sedz05good}, but for the bad-fit sample.
}
 \label{sedz05bad}
\end{figure*}

\begin{figure*}
 \resizebox{19cm}{!}{\includegraphics{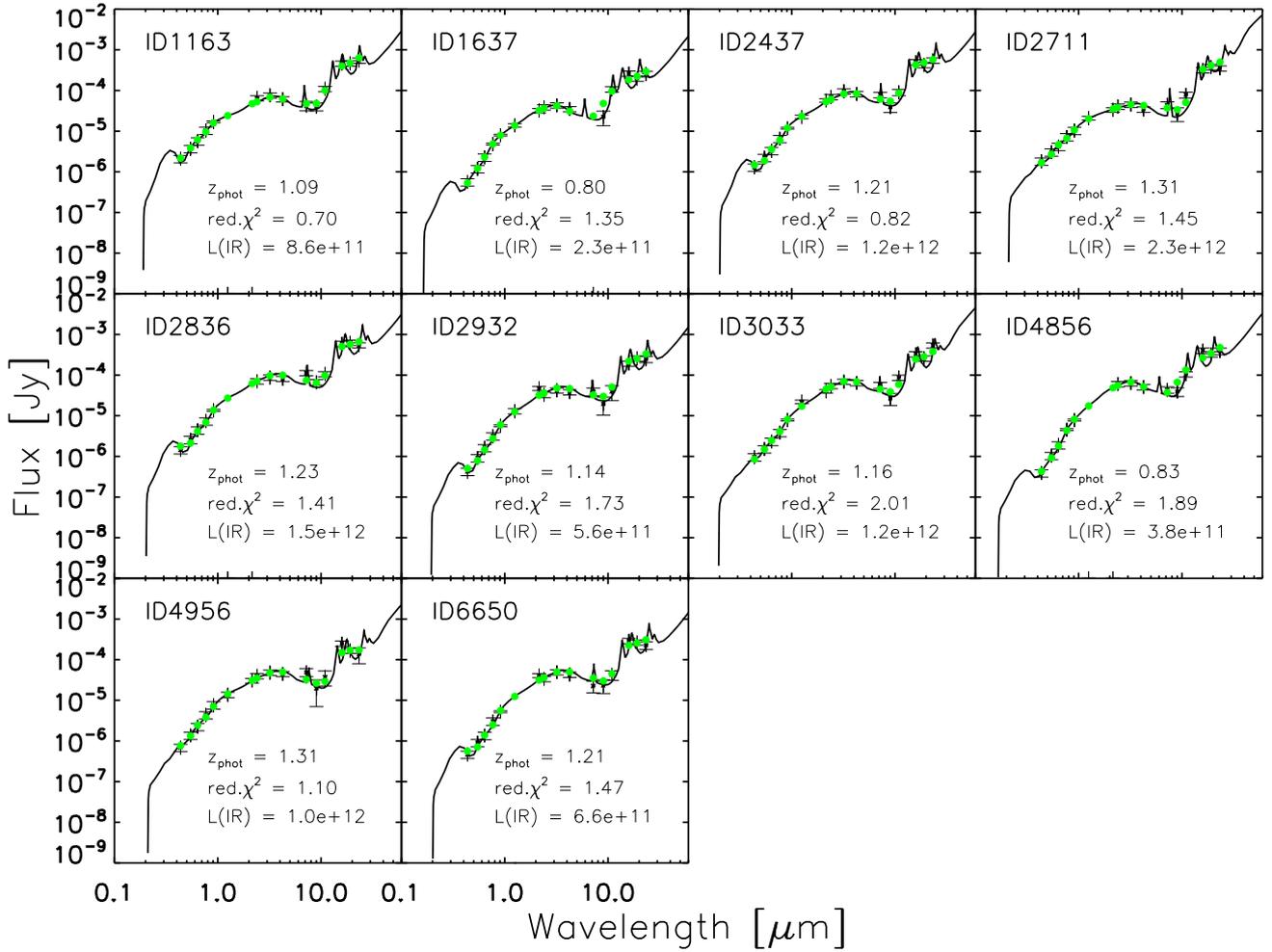}}
 \caption{Same as Figure \ref{sedz05good}, but for PAH-selected galaxies at $z\sim 1$ with acceptable fits.
}
 \label{sedz1good}
\end{figure*}

\begin{figure*}
  \resizebox{19cm}{!}{\includegraphics{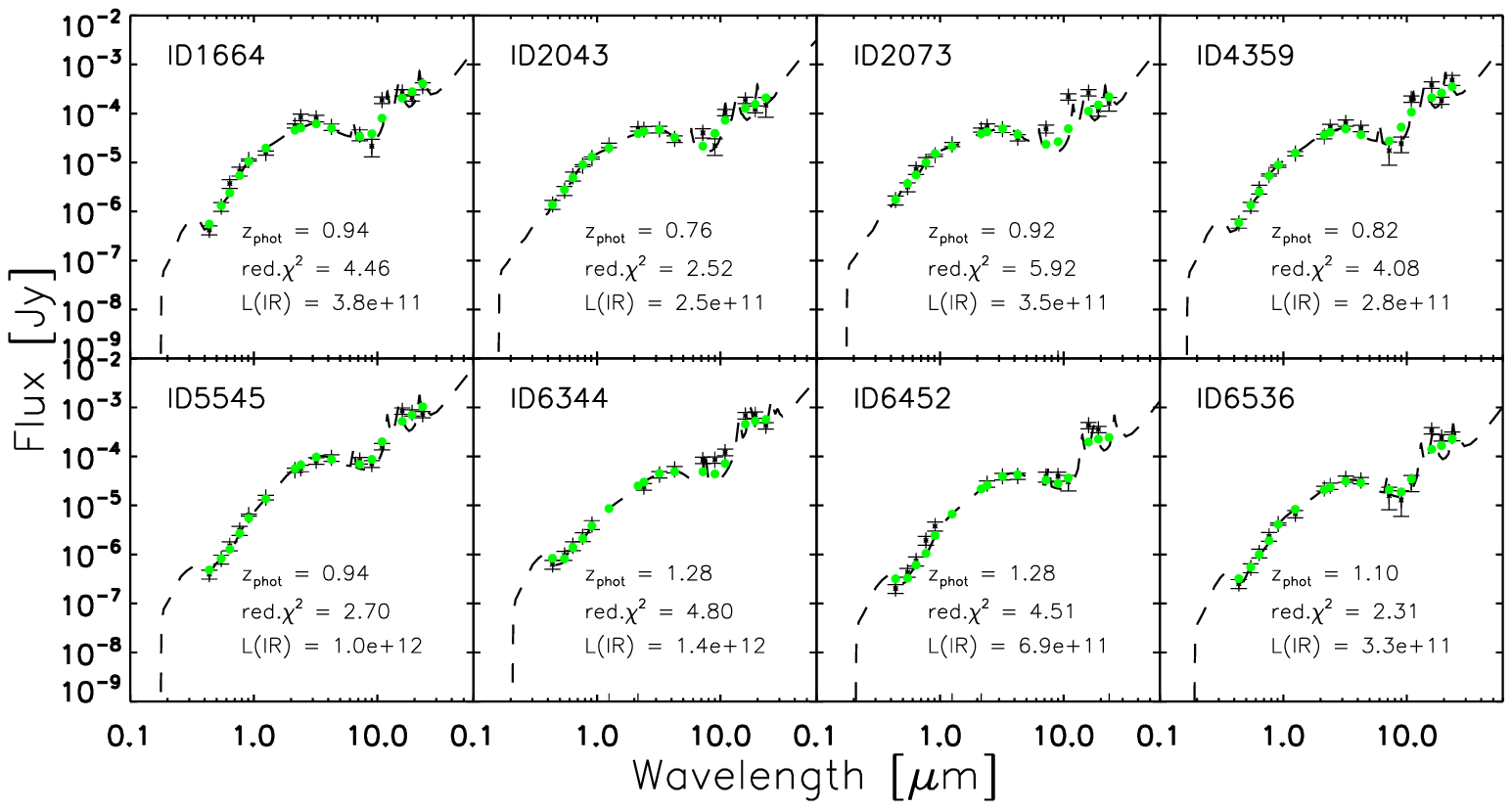}}
 \caption{Same as Figure \ref{sedz1good}, but for the bad-fit sample.
}
 \label{sedz1bad}
\end{figure*}

Here we show the results of the SED fitting analysis of PAH-selected galaxies in Figure \ref{sedz05good} -- \ref{sedz1bad}, and present the AKARI photometry in Table A.1.  
One source (ID4475) has problematic photometry in the S-cam images, due to a nearby bright source, and therefore is not included in the figure.

\begin{landscape}

\begin{table}
\caption{{\it AKARI} coordinates and flux densities of PAH-selexted galaxies in the S-cam field}
\tiny
\tiny
\begin{tabular}{lccccccccccccccccccccc}
\hline
ID & Name & RA & Dec & 
\multicolumn{2}{c}{${2.4\,\mu{\rm m}}$}& 
\multicolumn{2}{c}{${3.2\,\mu{\rm m}}$} & 
\multicolumn{2}{c}{${4.1\,\mu{\rm m}}$} & 
\multicolumn{2}{c}{${7.0\,\mu{\rm m}}$} & 
\multicolumn{2}{c}{${9.0\,\mu{\rm m}}$} & 
\multicolumn{2}{c}{${11\,\mu{\rm m}}$}  &
\multicolumn{2}{c}{${15\,\mu{\rm m}}$}  &
\multicolumn{2}{c}{${18\,\mu{\rm m}}$}  &
\multicolumn{2}{c}{${24\,\mu{\rm m}}$}  \\
\\
& &  \multicolumn{2}{c}{[J2000]}  &
$f_\nu$ & $\Delta f_\nu$ & 
$f_\nu$ & $\Delta f_\nu$ & 
$f_\nu$ & $\Delta f_\nu$ & 
$f_\nu$ & $\Delta f_\nu$ & 
$f_\nu$ & $\Delta f_\nu$ & 
$f_\nu$ & $\Delta f_\nu$ & 
$f_\nu$ & $\Delta f_\nu$ & 
$f_\nu$ & $\Delta f_\nu$ & 
$f_\nu$ & $\Delta f_\nu$ \\
& & & &
$\mu$Jy & $\mu$Jy & 
$\mu$Jy & $\mu$Jy & 
$\mu$Jy & $\mu$Jy & 
$\mu$Jy & $\mu$Jy & 
$\mu$Jy & $\mu$Jy & 
$\mu$Jy & $\mu$Jy & 
$\mu$Jy & $\mu$Jy & 
$\mu$Jy & $\mu$Jy & 
$\mu$Jy & $\mu$Jy \\
(1) & (2) & (3) & (4)  & (5) & (6) & (7) & (8) &
(9) & (10) & (11) & (12) & (13) & (14) & (15) &
(16) & (17) & (18) & (19) & (20) & (21) & (22)\\
\hline
\\
\multicolumn{21}{c}{PAH-selected galaxies at $z\sim 1$}  \\
\hline
  1163 & J175331.44+662050.3 & 268.38102 &  66.34731 &    64.8 &     2.2 &    75.7 &     2.0 &    61.1 &     1.7 &    42.6 &    11.1 &    40.3 &     8.9 &   110.0 &    13.9 &   461.0 &    24.0 &   466.0 &    22.1 &   545.0 &    51.4  \\
  1637 & J175611.19+662419.4 & 269.04663 &  66.40540 &    38.8 &     2.2 &    43.1 &     1.8 &    35.1 &     1.5 &     3.6 &     6.5 &    22.4 &     8.8 &   107.0 &    14.3 &   262.0 &    18.7 &   198.0 &    18.8 &   254.0 &    41.2  \\
  1664 & J175342.76+662421.9 & 268.42817 &  66.40611 &    84.0 &     3.9 &    79.3 &     2.5 &    53.0 &     1.7 &    37.0 &     9.2 &    21.4 &     8.4 &   188.0 &    13.2 &   268.0 &    23.1 &   210.0 &    21.8 &   363.0 &    59.8  \\
  2043 & J175323.76+662623.2 & 268.34902 &  66.43978 &    47.0 &     2.8 &    47.3 &     2.3 &    30.3 &     1.6 &    40.1 &     8.6 &    22.2 &     8.3 &   105.0 &    12.8 &   187.0 &    20.0 &   122.0 &    19.4 &   148.0 &    64.7  \\
  2073 & J175636.92+662653.6 & 269.15387 &  66.44824 &    53.3 &     2.3 &    47.9 &     2.2 &    32.2 &     1.9 &    48.8 &     8.9 &     9.9 &    12.9 &   220.0 &    20.4 &   266.0 &    25.0 &   115.0 &    27.5 &   162.0 &    50.6  \\
  2437 & J175322.46+662905.7 & 268.34360 &  66.48493 &    69.3 &     3.7 &    95.4 &     3.2 &    82.0 &     2.2 &    77.9 &    10.7 &    38.9 &    10.2 &    90.0 &    15.6 &   499.0 &    25.2 &   426.0 &    22.6 &   547.0 &    61.0  \\
  2711 & J175346.89+663038.2 & 268.44540 &  66.51062 &    40.3 &     1.9 &    45.0 &     1.6 &    34.2 &     1.3 &    44.1 &     8.5 &    27.3 &    10.1 &    77.9 &    13.3 &   318.0 &    23.4 &   378.0 &    22.4 &   351.0 &    50.4  \\
  2836 & J175726.41+663114.3 & 269.36005 &  66.52066 &    70.8 &     2.5 &    87.4 &     2.3 &    80.9 &     2.0 &   113.0 &     8.7 &    67.0 &    10.2 &   106.0 &    14.2 &   588.0 &    32.3 &   627.0 &    30.6 &   545.0 &    72.8  \\
  2932 & J175336.29+663153.2 & 268.40124 &  66.53146 &    33.0 &     1.8 &    42.1 &     1.8 &    38.2 &     1.4 &    39.6 &     8.7 &    19.4 &     9.1 &    39.2 &    15.5 &   192.0 &    20.2 &   256.0 &    17.6 &   246.0 &    43.7  \\
  3033 & J175458.78+663227.3 & 268.74493 &  66.54094 &    54.3 &     2.9 &    69.0 &     2.3 &    64.1 &     1.9 &    56.1 &     9.6 &    26.4 &     8.7 &    86.7 &    12.8 &   327.0 &    23.1 &   257.0 &    19.5 &   539.0 &    49.5  \\
  4359 & J175629.33+664530.6 & 269.12221 &  66.75850 &    52.4 &     2.7 &    64.0 &     2.2 &    49.7 &     1.7 &    17.5 &     8.7 &    24.3 &     8.5 &   198.0 &    12.5 &   383.0 &    33.5 &   185.0 &    30.5 &   482.0 &   111.0  \\
  4856 & J175731.87+664245.6 & 269.38281 &  66.71269 &    60.3 &     2.0 &    63.5 &     1.8 &    50.9 &     1.5 &    43.4 &     8.0 &    39.2 &     9.4 &   105.0 &    11.1 &   400.0 &    25.7 &   326.0 &    25.4 &   387.0 &    72.1  \\
  4956 & J175412.75+664151.2 & 268.55316 &  66.69757 &    39.8 &     3.5 &    48.1 &     2.3 &    44.6 &     2.0 &    49.2 &    11.3 &    19.4 &    12.4 &    37.6 &    15.0 &   250.0 &    25.1 &   168.0 &    25.3 &   137.0 &    57.5  \\
  5545 & J175328.41+663813.9 & 268.36838 &  66.63721 &    57.5 &     2.4 &    81.8 &     2.2 &    93.9 &     2.0 &    82.8 &     8.4 &    70.8 &    11.1 &   160.0 &    16.0 &   853.0 &    27.0 &   780.0 &    24.6 &   725.0 &    54.4  \\
  6344 & J175416.38+665332.4 & 268.56829 &  66.89236 &    24.2 &     2.3 &    42.7 &     2.1 &    54.3 &     1.8 &    84.3 &     9.0 &    85.1 &     8.0 &   122.0 &    12.8 &   685.0 &    26.6 &   709.0 &    26.9 &   425.0 &    60.5  \\
  6452 & J175732.98+665210.6 & 269.38742 &  66.86962 &    27.7 &     1.9 &    39.2 &     1.7 &    40.0 &     1.5 &    39.0 &     9.0 &    39.7 &     8.2 &    30.4 &    10.7 &   431.0 &    34.1 &   358.0 &    37.1 &    83.3 &    83.7  \\
  6536 & J175456.40+665137.9 & 268.73504 &  66.86055 &    25.0 &     1.7 &    34.8 &     1.6 &    32.4 &     1.4 &    15.9 &     7.7 &    13.0 &     7.0 &    30.0 &    10.8 &   339.0 &    26.0 &   243.0 &    27.2 &   263.0 &    54.7  \\
  6650 & J175433.94+665123.1 & 268.64142 &  66.85642 &    33.6 &     2.0 &    49.3 &     1.7 &    43.1 &     1.5 &    22.5 &     7.3 &    23.3 &     8.6 &    42.1 &    11.3 &   290.0 &    23.9 &   267.0 &    19.1 &   228.0 &    48.9  \\

\hline
\\
\multicolumn{21}{c}{PAH-selected galaxies at $z\sim 0.5$}  \\
\hline
  1174 & J175731.27+662050.0 & 269.38033 &  66.34723 &   116.0 &     3.3 &   112.0 &     2.5 &    79.1 &     1.9 &    49.4 &    10.4 &   189.0 &    13.0 &   416.0 &    17.8 &   514.0 &    30.4 &   538.0 &    33.9 &   620.0 &    60.7  \\
  1270 & J175526.98+662128.2 & 268.86244 &  66.35784 &    80.4 &     2.5 &    67.1 &     2.0 &    47.3 &     1.5 &    22.9 &     9.2 &   151.0 &    11.3 &   192.0 &    17.9 &   303.0 &    25.3 &   224.0 &    26.0 &   396.0 &    56.3  \\
  1321 & J175700.71+662141.3 & 269.25297 &  66.36148 &   142.0 &     3.3 &   112.0 &     2.5 &    79.7 &     1.9 &    58.4 &     9.0 &   328.0 &    12.3 &   501.0 &    18.9 &   485.0 &    24.6 &   594.0 &    23.6 &   841.0 &    56.7  \\
  1381 & J175650.63+662213.2 & 269.21097 &  66.37034 &    39.3 &     2.0 &    44.6 &     1.9 &    30.0 &     1.3 &    48.1 &     8.7 &   252.0 &    10.7 &   407.0 &    16.8 &   315.0 &    19.9 &   255.0 &    23.0 &   766.0 &    56.9  \\
  1451 & J175559.22+662245.6 & 268.99678 &  66.37935 &    85.1 &     2.6 &    70.8 &     2.1 &    52.9 &     1.7 &    23.3 &     8.1 &   102.0 &     8.9 &   226.0 &    14.3 &   262.0 &    23.8 &   227.0 &    20.2 &   186.0 &    42.2  \\
  1569 & J175600.05+662334.6 & 269.00021 &  66.39296 &    62.3 &     2.4 &    69.5 &     2.1 &    48.7 &     1.4 &    15.7 &     8.2 &   134.0 &     9.4 &   184.0 &    15.3 &   441.0 &    26.0 &   354.0 &    22.5 &   388.0 &    48.4  \\
  1642 & J175345.99+662357.8 & 268.44166 &  66.39940 &    48.4 &     3.7 &    55.4 &     3.5 &    39.2 &     2.3 &    35.4 &    10.4 &    93.8 &    10.1 &   306.0 &    15.6 &   325.0 &    26.9 &   406.0 &    25.0 &   403.0 &    56.1  \\
  1681 & J175640.43+662403.8 & 269.16847 &  66.40106 &    58.4 &     2.4 &    47.2 &     2.0 &    37.1 &     1.6 &    31.3 &     7.5 &   171.0 &     9.7 &   328.0 &    15.9 &   318.0 &    20.5 &   387.0 &    19.8 &   471.0 &    46.4  \\
  1726 & J175629.54+662444.9 & 269.12310 &  66.41249 &    81.9 &     2.5 &    64.9 &     2.0 &    42.2 &     1.5 &    27.4 &     7.9 &   175.0 &    13.1 &   313.0 &    14.8 &   236.0 &    20.9 &   329.0 &    19.7 &   238.0 &    49.0  \\
  1776 & J175631.43+662447.1 & 269.13098 &  66.41309 &   177.0 &     3.5 &   143.0 &     2.7 &   105.0 &     2.1 &    80.0 &     9.0 &   411.0 &    13.7 &   738.0 &    18.1 &   637.0 &    26.7 &   738.0 &    24.6 &   914.0 &    53.9  \\
  1901 & J175630.45+662550.4 & 269.12690 &  66.43068 &    55.6 &     2.1 &    42.7 &     1.6 &    30.0 &     1.2 &    13.0 &     8.3 &    94.5 &    14.1 &   237.0 &    20.3 &   147.0 &    19.7 &   143.0 &    23.9 &   222.0 &    50.2  \\
  1908 & J175518.76+662529.8 & 268.82820 &  66.42497 &   391.0 &     5.0 &   325.0 &     3.9 &   221.0 &     2.9 &   188.0 &    11.3 &   888.0 &    15.0 &  1540.0 &    21.9 &  1320.0 &    29.0 &  1370.0 &    26.3 &  1340.0 &    56.4  \\
  1934 & J175427.51+662545.6 & 268.61465 &  66.42934 &    87.7 &     2.5 &    67.0 &     2.1 &    57.8 &     1.7 &    18.2 &    10.6 &   329.0 &    12.6 &   543.0 &    17.5 &   414.0 &    23.7 &   518.0 &    23.6 &   520.0 &    48.0  \\
  2048 & J175415.43+662625.6 & 268.56430 &  66.44046 &   187.0 &     3.4 &   177.0 &     3.0 &   128.0 &     2.3 &    81.1 &     9.8 &   387.0 &    11.7 &   659.0 &    17.5 &   553.0 &    22.6 &   502.0 &    22.7 &   827.0 &    53.9  \\
  2111 & J175537.50+662702.4 & 268.90628 &  66.45067 &   133.0 &     2.9 &   105.0 &     2.3 &    73.7 &     1.7 &    72.6 &     9.0 &   276.0 &    12.1 &   722.0 &    18.1 &   581.0 &    25.0 &   562.0 &    24.1 &   437.0 &    48.2  \\
  2140 & J175641.58+662704.5 & 269.17325 &  66.45126 &   151.0 &     3.4 &   132.0 &     3.0 &   101.0 &     2.3 &   124.0 &     9.4 &   580.0 &    16.4 &   996.0 &    21.2 &   881.0 &    27.0 &  1040.0 &    27.0 &  1090.0 &    60.2  \\
  2256 & J175652.48+662736.3 & 269.21869 &  66.46011 &   200.0 &     3.6 &   181.0 &     2.9 &   121.0 &     2.2 &   139.0 &     9.1 &   778.0 &    20.0 &  1260.0 &    22.4 &  1030.0 &    29.2 &  1080.0 &    26.1 &   998.0 &    56.8  \\
  2626 & J175721.13+663009.3 & 269.33805 &  66.50260 &    37.8 &     1.8 &    35.0 &     1.4 &    26.2 &     1.2 &     2.3 &     6.7 &    94.8 &    10.9 &   143.0 &    14.7 &   229.0 &    27.7 &   236.0 &    29.4 &   201.0 &    70.6  \\
  2633 & J175352.59+662951.3 & 268.46913 &  66.49761 &    81.7 &     2.4 &    66.6 &     1.9 &    48.0 &     1.5 &    27.1 &     8.3 &   149.0 &    10.0 &   290.0 &    17.3 &   154.0 &    25.4 &   220.0 &    22.8 &   130.0 &    48.0  \\
  2675 & J175325.50+662957.2 & 268.35625 &  66.49923 &    69.9 &     2.2 &    63.7 &     1.8 &    38.9 &     1.3 &    28.1 &     8.6 &    85.0 &     8.4 &   235.0 &    13.7 &   414.0 &    25.1 &   291.0 &    22.5 &   453.0 &    53.9  \\
  2734 & J175723.08+663038.5 & 269.34619 &  66.51071 &    85.7 &     2.6 &    68.1 &     2.1 &    50.1 &     1.7 &    49.6 &     8.4 &   218.0 &    12.9 &   409.0 &    15.8 &   321.0 &    27.6 &   376.0 &    29.4 &   238.0 &    69.1  \\
  2833 & J175346.60+663101.8 & 268.44418 &  66.51717 &    43.2 &     2.0 &    34.0 &     1.6 &    27.3 &     1.4 &    11.7 &     8.1 &    65.8 &     8.7 &   184.0 &    15.4 &   190.0 &    22.1 &   196.0 &    20.8 &   210.0 &    45.9  \\
  2960 & J175327.78+663138.8 & 268.36579 &  66.52745 &    78.2 &     2.5 &    69.8 &     2.0 &    44.3 &     1.5 &    13.9 &     8.5 &    79.4 &    12.1 &   224.0 &    16.0 &   151.0 &    21.6 &   168.0 &    21.0 &   146.0 &    44.1  \\
\hline

\end{tabular}

Notes.--- Galaxies with ambiguous optical identification are excluded. Column~(1): AKARI mid-IR source ID. Columns~(2): Source Name. Columns~(3) and (4): AKARI $N2$-band J2000.0 RA and Dec. Columns~(5) through (22): IRC flux densities and errors. 
\end{table}

\end{landscape}

\begin{landscape}

\addtocounter{table}{-1}
\begin{table}
\caption{-- {\it continued}}
\tiny
\tiny
\begin{tabular}{lccccccccccccccccccccc}
\hline
ID & Name & RA & Dec & 
\multicolumn{2}{c}{${2.4\,\mu{\rm m}}$}& 
\multicolumn{2}{c}{${3.2\,\mu{\rm m}}$} & 
\multicolumn{2}{c}{${4.1\,\mu{\rm m}}$} & 
\multicolumn{2}{c}{${7.0\,\mu{\rm m}}$} & 
\multicolumn{2}{c}{${9.0\,\mu{\rm m}}$} & 
\multicolumn{2}{c}{${11\,\mu{\rm m}}$}  &
\multicolumn{2}{c}{${15\,\mu{\rm m}}$}  &
\multicolumn{2}{c}{${18\,\mu{\rm m}}$}  &
\multicolumn{2}{c}{${24\,\mu{\rm m}}$}  \\
\\
& &  \multicolumn{2}{c}{[J2000]}  &
$f_\nu$ & $\Delta f_\nu$ & 
$f_\nu$ & $\Delta f_\nu$ & 
$f_\nu$ & $\Delta f_\nu$ & 
$f_\nu$ & $\Delta f_\nu$ & 
$f_\nu$ & $\Delta f_\nu$ & 
$f_\nu$ & $\Delta f_\nu$ & 
$f_\nu$ & $\Delta f_\nu$ & 
$f_\nu$ & $\Delta f_\nu$ & 
$f_\nu$ & $\Delta f_\nu$ \\
& & & &
$\mu$Jy & $\mu$Jy & 
$\mu$Jy & $\mu$Jy & 
$\mu$Jy & $\mu$Jy & 
$\mu$Jy & $\mu$Jy & 
$\mu$Jy & $\mu$Jy & 
$\mu$Jy & $\mu$Jy & 
$\mu$Jy & $\mu$Jy & 
$\mu$Jy & $\mu$Jy & 
$\mu$Jy & $\mu$Jy \\
(1) & (2) & (3) & (4)  & (5) & (6) & (7) & (8) &
(9) & (10) & (11) & (12) & (13) & (14) & (15) &
(16) & (17) & (18) & (19) & (20) & (21) & (22) \\
\hline
  3110 & J175331.90+663231.9 & 268.38295 &  66.54220 &    65.1 &     2.3 &    61.9 &     1.9 &    40.7 &     1.5 &    13.4 &     8.0 &    63.3 &    10.7 &   230.0 &    19.0 &   149.0 &    20.4 &   184.0 &    22.8 &   218.0 &    46.7  \\
  3387 & J175345.13+663405.1 & 268.43805 &  66.56809 &    52.5 &     2.5 &    44.4 &     2.2 &    31.3 &     1.7 &     9.7 &     8.5 &    93.6 &    11.8 &   250.0 &    19.4 &   174.0 &    24.0 &   134.0 &    20.3 &    33.7 &    45.4  \\
  3953 & J175456.90+664818.7 & 268.73711 &  66.80520 &    93.2 &     2.7 &    84.9 &     2.3 &    62.1 &     1.8 &    11.5 &     8.2 &   180.0 &    10.4 &   311.0 &    14.1 &   305.0 &    27.3 &   294.0 &    21.9 &   256.0 &    50.3  \\
  4359 & J175629.33+664530.6 & 269.12221 &  66.75850 &    52.4 &     2.7 &    64.0 &     2.2 &    49.7 &     1.7 &    17.5 &     8.7 &    24.3 &     8.5 &   198.0 &    12.5 &   383.0 &    33.5 &   185.0 &    30.5 &   482.0 &   111.0  \\
  4438 & J175319.00+664419.7 & 268.32918 &  66.73881 &   118.0 &     2.9 &   103.0 &     2.3 &    68.0 &     1.7 &    20.2 &     7.2 &   150.0 &    10.3 &   215.0 &    13.4 &   218.0 &    20.9 &   245.0 &    19.1 &   308.0 &    49.1  \\
  4454 & J175335.34+664522.4 & 268.39727 &  66.75624 &    72.0 &     2.7 &    67.1 &     2.1 &    47.9 &     1.7 &    20.8 &     7.7 &    89.8 &    11.1 &   203.0 &    14.0 &   251.0 &    21.9 &   270.0 &    19.2 &   268.0 &    48.1  \\
  4653 & J175645.64+664334.9 & 269.19020 &  66.72637 &    82.1 &     3.2 &    74.2 &     2.4 &    51.6 &     2.2 &    37.6 &     8.1 &   296.0 &    10.1 &   463.0 &    12.9 &   333.0 &    27.8 &   347.0 &    23.3 &   465.0 &    64.4  \\
  4802 & J175328.00+664228.0 & 268.36667 &  66.70779 &    66.5 &     2.3 &    67.2 &     2.2 &    43.9 &     2.0 &     8.9 &     6.6 &    60.5 &     9.2 &   200.0 &    11.2 &   245.0 &    19.8 &   132.0 &    17.7 &   143.0 &    44.7  \\
  5339 & J175651.89+664003.0 & 269.21623 &  66.66751 &    50.0 &     2.4 &    45.8 &     2.2 &    28.8 &     2.1 &     7.8 &     9.7 &    60.2 &    10.5 &   217.0 &    13.2 &   209.0 &    32.0 &   216.0 &    29.5 &   247.0 &    62.9  \\
  5424 & J175534.22+663918.4 & 268.89262 &  66.65512 &    53.7 &     2.0 &    65.1 &     1.9 &    50.2 &     1.4 &    42.7 &    11.4 &   136.0 &    12.7 &   538.0 &    19.1 &   487.0 &    35.9 &   522.0 &    47.6 &  1010.0 &    96.6  \\
  5776 & J175328.61+663701.4 & 268.36922 &  66.61707 &    53.6 &     2.0 &    48.8 &     1.8 &    31.0 &     1.4 &     3.2 &     7.0 &    75.2 &    10.8 &   162.0 &    16.3 &   200.0 &    19.8 &   195.0 &    18.2 &   250.0 &    47.6  \\
  6363 & J175701.56+665331.4 & 269.25654 &  66.89206 &    54.1 &     2.4 &    61.1 &     2.3 &    39.0 &     1.7 &     4.4 &     7.4 &    86.4 &     9.8 &   197.0 &    13.2 &   270.0 &    25.6 &   278.0 &    30.6 &   507.0 &    86.5  \\
  6393 & J175402.88+665222.2 & 268.51203 &  66.87284 &   257.0 &     4.5 &   229.0 &     3.3 &   174.0 &     2.8 &   176.0 &    10.2 &   828.0 &    11.9 &  1420.0 &    17.9 &  1190.0 &    29.8 &  1400.0 &    28.7 &  1780.0 &    62.0  \\
  6427 & J175430.88+665249.1 & 268.62869 &  66.88033 &   141.0 &     3.1 &   132.0 &     2.7 &    89.1 &     2.1 &    65.5 &     7.9 &   342.0 &     9.4 &   630.0 &    14.9 &   636.0 &    26.4 &   537.0 &    22.1 &   653.0 &    50.6  \\
  6449 & J175542.96+665248.9 & 268.92901 &  66.88027 &    37.2 &     1.9 &    39.3 &     1.7 &    27.6 &     1.4 &     7.5 &     7.8 &    72.5 &     8.4 &   213.0 &    12.5 &   232.0 &    28.0 &   274.0 &    23.5 &   426.0 &    57.9  \\
\hline

\end{tabular}

Notes.--- Column~(1): AKARI mid-IR source ID. Columns~(2): Source Name. Columns~(3) and (4): AKARI $N2$-band 
J2000.0 RA and Dec. Columns~(5) through (22): IRC flux densities and errors. 
\end{table}

\end{landscape}

\end{appendix}

\end{document}